\documentclass[reprint,aps,nofootinbib,preprintnumbers,amsmath,amssymb,11pt]{revtex4}
\usepackage{epsfig}
\usepackage{amsfonts}
\usepackage{amsmath}
\usepackage[T1]{fontenc}
\usepackage[latin2]{inputenc}
\def\bea{\begin{eqnarray}}
\def\eea{\end{eqnarray}}
\def\bq{\begin{quote}}
\def\eq{\end{quote}}

\def\gappeq{\mathrel{\rlap
{\raise.5ex\hbox{$>$}}
{\lower.5ex\hbox{$\sim$}}}}
\def\lappeq{\mathrel{\rlap{\raise.5ex\hbox{$<$}}
{\lower.5ex\hbox{$\sim$}}}}
\def\simlt{\stackrel{<}{{}_\sim}}
\def\simgt{\stackrel{>}{{}_\sim}}
\newcommand{\beq}{\begin{equation}}
\newcommand{\eeq}{\end{equation}}
\newcommand{\p}{\partial}
\newcommand{\half}{\frac{1}{2}}

\newcounter{mnotecount}[section]

\begin{document}

\preprint{IFT-10-14}
\preprint{DAMTP-2010-80}
\preprint{MIFPA-10-43}

\title{\bf Strongly Coupled Perturbations in Two-Field Inflationary Models}

\author{Sera Cremonini$ ^{\,\clubsuit,\spadesuit}\,$}
\email{sera@physics.tamu.edu,  S.Cremonini@damtp.cam.ac.uk}
\affiliation{$ ^\clubsuit$ Centre for Theoretical Cosmology, DAMTP, CMS,\\
\it University of Cambridge, Wilberforce Road, Cambridge, CB3 0WA, UK \\
%[.5em]
\it $ ^\spadesuit$ George and Cynthia Mitchell Institute for Fundamental Physics and Astronomy\\
\it Texas A\&M University, College Station, TX 77843--4242, USA}

\author{Zygmunt Lalak}
\email{lalak@fuw.edu.pl}
\affiliation{Institute of Theoretical Physics, Warsaw University, ul.\ Ho\.za 69, 00-681 Warsaw, Poland}

\author{Krzysztof Turzy\'nski}
\email{turzyn@fuw.edu.pl}
\affiliation{Institute of Theoretical Physics, Warsaw University, ul.\ Ho\.za 69, 00-681 Warsaw, Poland}

\begin{abstract}
We study
models of inflation with two scalar fields and non-canonical kinetic
terms, focusing on the case in which the curvature and isocurvature perturbations are strongly coupled to each other.
In the regime where a heavy mode can be identified and integrated out, we
clarify the passage from the full two-field model to an effectively single-field description.
However, the strong coupling sets a new scale in the system, and affects the evolution of the perturbations
as well as the beginning of the regime of validity of the effective field theory.
In particular, the predictions of the model are sensitive to the relative hierarchy between the coupling and the mass of the heavy mode.
As a result, observables are not given unambiguously in terms of the parameters of an effectively single field model with non-trivial sound speed.
Finally, the requirement that the sound horizon crossing occurs within the regime of validity of the effective theory leads to a
lower bound on the sound speed.
Our analysis is done in an extremely simple toy model of slow-roll inflation, which is chosen for its
tractability, but is non-trivial enough to illustrate the richness of the dynamics in non-canonical multi-field models.
\end{abstract}

\maketitle
\newpage

\tableofcontents

%%%%
%%%%
\section{Introduction}
\label{Intro}
%%%%
%%%%

Over the years inflation has proven to be remarkably successful at describing cosmological evolution.
However, it is still unclear how it might arise within the framework of a more fundamental theory.
For example, inflationary model building in the context of string theory has been faced with many challenges
(see, {\em e.g.}, \cite{Quevedo:2002xw,McAllister:2007bg,Baumann:2009ni} for comprehensive reviews).
Nonetheless, one can still try to ask whether any of the generic ingredients of string theory
-- the presence of extra dimensions, new symmetries, the large number of light scalar fields --
might lead to distinct predictions on low-energy effective field theories, and in particular leave imprints
on inflationary dynamics.

It is well known that string theory compactifications come with a variety
of scalar moduli fields which parametrize the geometry of the extra dimensions, as well as the string coupling.
The fact that such light fields can have problematic cosmological consequences,
such as the late entropy production \cite{Coughlan:1984yk}, the
overclosure problem \cite{Holman:1984yj} or the overshoot problem \cite{Brustein:1992nk},
has driven a large
effort to find mechanisms to stabilize them.
Also, having models with moduli stabilized is the starting point for quasi-realistic
phenomenology, and allows for the computation of relevant scales (the string scale, the gravitino mass,
etc) from first principles \cite{Dine:1985vv}.
Recent years have seen much progress in this direction, with the emergence of concrete moduli-fixing
techniques (for reviews see, {\em e.g.}, \cite{Giddings:2001yu,Silverstein:2004id,Douglas:2006es}).
An obvious application of moduli potentials is to the realm of inflation -- once some of the light scalars acquire a potential,
it is natural to ask whether the potential may support inflationary conditions
and whether the predictions for the inflationary observables agree with the measurements.

At present, the most relevant observables are the fluctuations of the
cosmic microwave background (CMB), accurately measured by
the WMAP satellite \cite{Komatsu:2008hk}.
Inflationary predictions for the detectable CMB fluctuations depend mainly on two
parameters:
the normalization of the power spectrum of the curvature
perturbations, $\mathcal{P}_\mathcal{R}$,
and the running of the spectrum of these perturbations, $n_s$.
In slow-roll single-field models, these quantities are linked to the inflationary potential $V$ by
the following relations (see {\em e.g.} \cite{mukhanov}):
\beq
\label{psf}
\mathcal{P}_\mathrm{sf} = \frac{V}{24\pi^2M_P^4\epsilon}\,,\qquad n_s=1-6\epsilon+2\eta\, ,
\eeq
where $\epsilon\equiv -\dot H/H^2$ and $\eta=V''/(3H^2)$.
The sensitivity of the PLANCK satellite \cite{planck}, currently taking data, and future missions may
lead to discovery of even more subtle features of the CMB, {\em e.g.} primordial tensor perturbations or
primordial non-gaussianities.

In this paper we would like to go beyond minimal models of inflation and
consider scenarios with more than one scalar field active during inflation,
focusing in particular on the role of non-canonical kinetic terms.
This type of setting is well-motivated by known examples of string compactifications,
in which the dynamics of the moduli $X^I$ is often governed by a non-trivial moduli space metric $G_{IJ}$.
In the lower-dimensional effective action, this gives rise to a kinetic term for the scalars of the form
$\mathcal{L}_{kin} = G_{IJ} \, \p_\mu X^I \p^\mu X^J$.
As long as the moduli-space metric is not flat, one is lead generically to non-canonical kinetic terms.
Such terms can enhance the coupling between the curvature perturbations
(measured in the CMB) and the isocurvature perturbations, resulting in interesting predictions
\cite{GrootNibbelink:2000vx,GrootNibbelink:2001qt,DiMarco:2002eb,vanTent:2003mn,DiMarco:2005nq,Lalak:2007vi,Vincent:2008ds,Peterson:2010np}.
For example, such a coupling can affect the evolution of the perturbations on super-Hubble scales,
{\em e.g.}\ modifying the running of the spectral index and leading to an enhancement of the redness
of the spectrum, as was shown in \cite{Cremonini:2010sv}.
Such terms are also relevant for non-gaussianities,
which are often enhanced in multi-field models with non-canonical kinetic terms, even if
only one of the fields drives inflation \cite{Chen:2009zp}.

With these motivations in mind, we wish to extend our study of models of two-field inflation
\cite{Cremonini:2010sv}, by
concentrating on the effects of a large coupling between the curvature and isocurvature perturbations.
This research was initiated by the authors of \cite{Tolley:2009fg}, who found
that the full evolution of the perturbations in such models, on sufficiently
large scales, may be effectively described by single-field models with a speed of sound
smaller than unity -- a non-decoupling effect of the heavy isocurvature mode integrated out.
We will extend these results in many aspects.
First, we will show that the presence of such a coupling leads to a suppression of
the perturbations well inside the Hubble radius, before the passage to the effective theory can be made.
Second, we will find that in the effective theories of \cite{Tolley:2009fg}
({\em i.e.}, assuming a sound horizon crossing within the regime of validity of the effective theory)
there exists a lower bound on the speed of sound.
We will also show that the passage to the effective theory (and in particular the beginning
of its regime of validity) is different if the above assumption is not satisfied
-- a direct result of the fact that the large coupling introduces a new scale in the system.
This will in turn affect the inflationary observables.
Last but not least, we will argue that for certain choices of the parameters
in our models, curvature perturbations develop a temporary instability around the Hubble
radius crossing, which can significantly enhance their amplitude.
All these effects can have important implications for the spectrum of the curvature perturbations.

The paper is organized as follows.
In Section \ref{Ingredients} we briefly introduce our very simplified inflationary model that
will serve as an illustration of the origin of possible new effects in the inflationary dynamics.
Section \ref{Predictions} contains a discussion of the evolution of the perturbations,
and Section \ref{GEL} a comparison to the dynamics observed in effective single-field models.
Since not all aspects of the two-field dynamics can be captured by such models,
we present a more complete catalogue of predictions in Section \ref{CAT}.
We summarize and discuss our results in Section \ref{Discussion}.
The Appendix contains detailed calculations of the results that we refer to in Section \ref{Predictions}.

%%%%
%%%%
\section{The Model}
\label{Ingredients}
%%%%
%%%%

Our main interest in this paper is in exploring the effects of a {large coupling} between the curvature and the isocurvature
perturbations,
due to a  non-canonical term.
We will consider two-field models described by an effective Lagrangian of the form
\beq
\label{ourL}
\mathcal{L}_\mathrm{eff} = R -\frac{1}{2}\, g^{\mu\nu} \partial_\mu\phi \, \partial_\nu\phi
- \frac{1}{2} \,e^{2\,b(\phi)} \, g^{\mu\nu} \partial_\mu\chi \, \partial_\nu\chi -V(\phi,\chi) \, ,
\eeq
with $\chi$ playing the role of the inflaton field, and $\phi$
coupling to $\chi$ through $b(\phi)$ in the non-canonical kinetic terms.
Following our previous analysis of the small coupling scenario \cite{Cremonini:2010sv},
we make several rather simplifying assumptions, and choose the ingredients of the model to isolate
new effects arising in the presence of strong non-canonicality.
Namely, we choose a trajectory in field space which corresponds to holding $\phi$ constant.
Thus, $\phi$ plays the role of a spectator field.
The non-trivial curvature of the field-space metric enables the isocurvature perturbations to affect
the curvature perturbation, even in the absence of a direct interaction term in the potential.
As we shall see in more detail later, the coupling between the curvature and isocurvature
perturbations is described by a dimensionless quantity
\beq
\xi \equiv M_P \sqrt{2\epsilon} \partial_\phi b\, .
\eeq
Motivated by typical string theory compactifications, we assume that $b$ is a linear function of $\phi$.
We also assume -- to make the analysis tractable -- that the potential takes the simple form
\beq
V(\phi,\chi) = V_0 \left[ 1+ \alpha \left(\frac{\phi-\phi_0}{M_P} \right)^2 + \beta \,\frac{\chi-\chi_0}{M_P} \right] \, .
\eeq
Since the spectator field $\phi$ is approximately constant, it is natural to expand the potential around
its minimum, and expect a quadratic dependence of this type. Also, the linear potential for the inflaton $\chi$
can be motivated by invoking an approximate shift symmetry, which is not spoiled by the non-canonicality,
since the latter is only a function of $\phi$.
In fact, there are string theory constructions \cite{msw,msw2}
which make use of axion monodromies and
approximate shift symmetries to construct linear inflaton potentials, and result in super-Planckian field excursions (see also \cite{fibre}).

While kinetic terms of the type of (\ref{ourL}) arise generically in many supergravity constructions,
the potential is an obvious simplification -- we are ignoring various corrections (for example,
corrections to the leading order dimensional reduction), and a proper analysis would have to include such effects.
One should also worry about whether it is natural to take the coupling $\xi$ between the curvature and isocurvature
perturbations to be large.
However, our approach here is rather phenomenological --
we assume that one can have enhanced non-canonicality, and study a toy model which serves as a
playground for better understanding the rich structure of multi-field dynamics.
There may be {generic} lessons to be learned by studying such a set-up, independently of its UV completion.
Our interest is in the evolution of the full two-moduli system in the presence of strongly coupled perturbations,
with particular emphasis on whether any qualitative signatures might be somehow lost in attempting to reduce the model to a single-field one.
The great simplification offered by our toy model is that the parameters entering the perturbation equations
have no implicit time dependence, making the analysis analytically tractable\footnote{We
checked numerically that this holds for the parameter choices considered here.}.

%%%%%%%%%%%%%%%%%%%%%%%%
%%%%%%%%%%%%%%%%%%%%%%%%
\section{Strongly Coupled Curvature and Isocurvature Perturbations}
\label{Predictions}
%%%%%%%%%%%%%%%%%%%%%%%%
%%%%%%%%%%%%%%%%%%%%%%%%

It is well-known that in multi-field inflationary models the evolution of the curvature and
isocurvature perturbations is {coupled on super-Hubble scales} if the inflationary
trajectory in field space is curved (see {\em e.g.}\ \cite{Gordon:2000hv} for the action expanded to the second order
in perturbations and \cite{Langlois:2008mn,Gao,Chen:2009zp} for the third-order action)
and/or the field-space metric itself has a nontrivial curvature (see {\em e.g.}\
\cite{GrootNibbelink:2000vx,GrootNibbelink:2001qt,DiMarco:2002eb,vanTent:2003mn,DiMarco:2005nq,Lalak:2007vi,Vincent:2008ds,Peterson:2010np}).
In this section we discuss the main features of the evolution of the coupled curvature and isocurvature perturbations,
for the simple setup we introduced in Section \ref{Ingredients}, leaving all details to the Appendix.
We also identify the regime in which the two-field dynamics can be described in terms
of an effectively single field theory.

It is particularly convenient to work with $u_\sigma=Q_\sigma/a$ and $u_s=\delta s/a$, where
$Q_\sigma$ and $\delta s$ are the Mukhanov-Sasaki variables associated with, respectively, the
perturbations along the field trajectory and orthogonal to it. These are often referred to as the instantaneous adiabatic
(or curvature) and entropy (or isocurvature) perturbations \cite{Gordon:2000hv}.
Assuming that the effects of the coupling $\xi$ dominate over the contribution from the potential
(with a possible exception of a large mass of the perturbation perpendicular to the inflationary trajectory,
described by a `slow-roll' parameter $\eta_{ss}\equiv V_{ss}/3H^2$),
the perturbations equations \cite{Lalak:2007vi} in conformal time $\tau$ become:
\beq
\label{eomgel0}
\left[ \left(\frac{\mathrm{d}^2}{\mathrm{d}\tau^2}+k^2-\frac{2}{\tau^2}\right)
+\left(\begin{array}{cc} 0 & \frac{2\xi}{\tau} \\ -\frac{2\xi}{\tau} & 0 \end{array}\right)
\frac{\mathrm{d}}{\mathrm{d}\tau}+
\left(\begin{array}{cc} 0  & -\frac{4\xi}{\tau^2} \\
-\frac{2\xi}{\tau^2} & \frac{1}{\tau^2}(3\eta_{ss}-2\xi^2) \end{array}\right)\right]
\left(\begin{array}{c}u_1 \\ u_2 \end{array}\right)=0 \, .
\eeq
They can be further rewritten along the lines of \cite{Lalak:2007vi} in a way that resembles more closely a standard harmonic oscillator,
and makes the early-time evolution of the modes particularly simple to study.
Introducing a new basis $\vec{\mathcal{U}}=R^{-1}\vec u$ for the perturbations, where $R$ is a time-dependent rotation matrix,
\beq
\label{Rmatrix}
R=\left(\begin{array}{cc}\cos(\xi\log(-k\tau))&-\sin(\xi\log(-k\tau)) \\ \sin(\xi\log(-k\tau))& \cos(\xi\log(-k\tau))\end{array}\right),
\eeq
the perturbation equations (\ref{eomgel0}) take the much simpler harmonic oscillator form
\beq
\label{btr2}
\vec{\mathcal{U}}''+\left(k^2-\frac{2}{\tau^2}+\frac{1}{\tau^2} R^T\mathcal{M}R\right)\vec{\mathcal{U}} = 0 \, ,
\eeq
where we have introduced an effective `{mass matrix}' given by:
\beq
\label{tq2orig}
\mathcal{M} = \left( \begin{array}{cc} \xi^2 & -3\xi \\ -3\xi & 3\eta_{ss}-\xi^2 \end{array}\right)\, .
\eeq
We note that, for the case of canonical kinetic terms, where $\xi =0$, the system (\ref{btr2}) reduces to that
studied by many authors ({\em e.g.} in \cite{Byrnes:2006fr} and more recently also in \cite{Berglund:2009uf}),
where the perturbation equations (in the slow-roll
approximation) were written in the analogous form\footnote{In the limit
$\xi\to0$ our $M_{IJ}$ reduces to $\mathrm{diag}(0,-\eta_{ss})$, with the large negative eigenvalue signaling the presence of a heavy field,
and the behavior of the modes is as described in \cite{Berglund:2009uf}.}:
\beq
\label{ByrnesWands}
\mathcal{U}_I''+\left(k^2-\frac{2}{\tau^2}\right)\mathcal{U}_I=\frac{3}{\tau^2} \sum_J M_{IJ} \, \mathcal{U}_J\, .
\eeq
By comparing (\ref{btr2}) and (\ref{ByrnesWands}) it is now evident that the non-canonicality encoded by $\xi$ not only affects
the eigenvalues of the interaction matrix $M_{IJ}$, but also adds to it a strong time-dependent rotation.

To highlight the possible hierarchy of masses in the effective mass matrix $\mathcal{M}$, it turns out to be
convenient to introduce a parameter $\nu$ defined by
\beq
\nu\xi^2\equiv 3\eta_{ss}-\xi^2 \, ,
\eeq
in terms of which (\ref{tq2orig}) takes the suggestive form:
\beq
\label{tq2}
\mathcal{M} = \left( \begin{array}{cc} \xi^2 & -3\xi \\ -3\xi & \nu\xi^2 \end{array}\right)\, .
\eeq
Note that, while for $\nu\approx 1$ its mass eigenvalues are nearly equal, $\lambda_{1,2}\approx\xi^2\pm3\xi$,
for large $\nu$ there is a clear hierarchy, $\lambda_2 \sim \nu \lambda_1 + {\cal O}(\xi^{-2})\gg \lambda_1$.
As we will see in Section \ref{GEL}, the $\nu\sim 1$ case -- no hierarchy of masses in the effective mass matrix -- is precisely
what is needed to achieve a very small sound speed in the effectively single-field description.

We now move on to discussing briefly the behavior of the solutions of the perturbation equations,
under some symplifying assumptions. We relegate all details to the Appendix.
Deep inside the Hubble radius, at early times $\tau\to-\infty$, the last term in (\ref{btr2}) can be neglected, and
the system reduces to that of two uncoupled harmonic oscillators, describing two independent perturbations,
$\mathcal{U}^{(i)}_I\sim\delta_{iI}e^{-\imath k\tau}$.
After Hubble radius crossing, at sufficiently late times $k|\tau|<|\xi|$, we can neglect the $k$-dependent terms in (\ref{eomgel0}).
We then find four asymptotic solutions, shown in the Appendix.
One of them describes a growing mode of an almost massless perturbation,
corresponding to the {curvature} perturbation.
Another one is a decaying mode of an almost massless perturbation.
The other two solutions correspond to positive- and negative-energy solutions for a massive mode with an effective mass squared
\beq
\label{meff1}
m_\mathrm{eff}^2=V_{ss}+2\xi^2H^2=(3\eta_{ss}+2\xi^2)H^2 = (\nu+3)\xi^2H^2\, ,
\eeq
which we can associate with the {isocurvature} mode.
In fact, we can characterize the evolution of the perturbations more accurately than by
describing only their asymptotic behavior.  To this end, we can assume a solution in the form of
a power series in $(k\tau)$, and find its coefficients by solving (\ref{eomgel0}) order by order.
This somewhat tedious calculation is carried out for the curvature perturbation in the Appendix.

The remarks above seem to contradict the general notion that on super-Hubble scales
the curvature and the isocurvature perturbations should be strongly coupled for $\xi\gg1$.
However, we have just seen that only the curvature mode remains frozen-in, apparently unaffected by the presence of the isocurvature mode.
%%%
It is therefore interesting to study the evolution of the perturbations in more detail.
In particular, we would like to determine the effective single-field theory which can be used to describe the evolution of the curvature
modes on large scales, and the regime of its validity.
We are particularly interested in the $\nu\sim 1$ case which, corresponding to a very small sound speed, is the most relevant
phenomenologically.

In order to study the effective single-field theory,
it is useful to recast (\ref{btr2}) in a slightly different form.
By choosing appropriate combinations $\mu_\pm$ of the eigenvalues of the effective mass matrix, it is possible to
isolate the rapidly oscillating terms in (\ref{btr2}), and rewrite the perturbation equations as:
\beq
\label{btr3m}
\vec{\mathcal{U}}^{\, \prime\prime} +\left[ \left( k^2+\frac{\mu_+^2-2}{\tau^2}\right) +\frac{\mu_-^2}{\tau^2} \left(\begin{array}{cc} -
\cos(2\xi \log(-k \tau)) & \sin(2\xi \log(-k \tau)) \\ \sin(2\xi \log(-k \tau)) & \cos(2\xi \log(-k \tau))\end{array}\right)\right]
\vec{\mathcal{U}} =0 \, .
\eeq
The precise expressions for $\mu_\pm$ are shown in Appendix. Here we simply note that
for $\nu \sim 1$ the two parameters can be approximated by:
\beq
\label{apprmupm}
\mu_+^2 \sim \xi^2 \, , \quad \quad \mu_-^2\sim \mathrm{max}\left( 3\xi,(\nu-1)  \frac{\xi^2}{2} \right) \, .
\eeq
Since we are working under the assumption of strong coupling, $\xi \gg 1$, our parameter choice $\nu \sim 1$ corresponds to $\mu_+\gg \mu_-$.
Thanks to this hierarchy, the effects of the rapid rotation are suppressed,
and (\ref{btr3m}) can be solved perturbatively.
Thus, as long as the $\mu_-$ term can be neglected, the solutions of  (\ref{btr3m}) consist of two
independently evolving modes\footnote{In the limit $\mu_-\to0$, the solutions can be written in terms of Bessel functions.
Here we give their approximate behavior, assuming that $\mu_+\gg 1$ and $k\tau \ll \mu_+$, and neglecting factors
$k^2$ and $1/\tau^2$ in the uncoupled equation.}
{of nearly equal masses} $\sim \mu_+ H$,
\beq
\label{leading}
\vec{\mathcal{U}}_{(0)}^{\,(a,b)}\sim \,\sqrt{-k\tau} \,e^{-\imath \mu_+\log(-k\tau)} \vec e_\pm \, ,
\quad
\quad \vec e_\pm=\left( \begin{array}{c} 1\\ \pm\imath \end{array}\right) \,.
\eeq
The leading solution (\ref{leading}) can be easily corrected, by plugging
$\vec{\mathcal{U}}_{(0)}$ into the $\mu_-$ term in (\ref{btr3m}), and treating it as a source.
Details of this iterative procedure are described in the Appendix.

What do we learn from this analysis?
First, in the leading order approximation in which the $\mu_-$ term is neglected,
one can {unambiguously} identify the mass eigenvalues, and finds that the two modes have the same behavior and
nearly equal masses. This tells us that there is no naturally heavy mode
which can be identified and integrated out, and therefore no effectively single-field description of the dynamics of the system.
Furthermore,
to ensure that our correction
to (\ref{leading}) is in fact small, and in particular smaller than any of
the terms dropped in the leading order approximation, we need $\mu_-^2/(k\tau)^2\ll 1$.
Thus, our approximations and perturbative procedure break down when $|k\tau|\sim {\cal O} (\mu_-)$.
Recalling from (\ref{apprmupm}) how $\mu_-$ was defined, we see that this break-down happens when
\beq
\label{EFTbegins}
-k\tau \sim  \mathrm{max}\{\sqrt\xi,\sqrt{\nu-1}\xi\} \, ,
\eeq
where we dropped numerical factors of order one.
Thus, we expect that the period during which both
fields decay as massive modes (with nearly equal masses) ends as soon as $-k\tau$ reaches the {\em larger} of the values
$\sim\sqrt{\xi}$ or $\sim\sqrt{\nu-1}\xi$.
This time marks the onset of the asymptotic behavior of the solutions
and the beginning
of the effective single-field description, as we describe in more detail next.

\section{The Passage to the Effective Single-Field Description}
\label{GEL}

The analysis of the super-Huble evolution in Section \ref{Predictions} revealed the existence of
a massless mode corresponding to the curvature perturbation and a massive mode describing the isocurvature perturbation.
If such a mass hierarchy is present, one can in principle integrate out the heavy degree of freedom,
and obtain an effectively single-field description.
In fact, the same inflationary model we are discussing here -- with a large coupling between
curvature and isocurvature perturbations -- was studied previously in \cite{Tolley:2009fg},
under the assumption that the isocurvature perturbations are heavy (compared to the Hubble scale).
In that regime, the authors of \cite{Tolley:2009fg} claimed that it is legitimate to consider an effective single-field
theory of inflation, whose equations of motion are very similar to ordinary models of inflation
-- with the exception that the propagation of the curvature modes is slowed down by interactions
with the isocurvature perturbations (hence the name, {\sl the gelaton scenario}).
This slow-down can be understood in terms of the presence of an effective speed of sound $c_s$
in the equation of motion of the curvature perturbations.
The  authors of \cite{Tolley:2009fg} concluded
that such models are capable of mimicking the dynamics of known single-fields models
with $c_s\neq 1$, of which Dirac-Born-Infeld (DBI) inflation \cite{dbisky} is a prime example and, in particular, that
large non-gaussianities can be generated in the gelaton scenario.

Here we would like to compare our full two-field dynamics to that of the gelaton model.
We are particularly interested in asking whether there are any (potentially significant) qualitative differences
between the original two-field model and the effective single-field description obtained by integrating out the heavy gelaton field.
However, as alluded to in Section \ref{Predictions} and discussed in much detail in the Appendix, the
identification of mass parameters in the equations of motion can be basis-dependent.
Therefore, a correct identification of the mass parameters of the various fields can be quite subtle,
especially when a rapid rotation links the bases natural for the early-time evolution
to those appropriate for the late-time evolution.
We note that this subtlety is an example of a more general feature discussed already in \cite{Burgess:2002ub}: the presence
of fast oscillations introduces an additional mass scale, which differs from the Hubble parameter or the field masses.
This mass scale in turn affects the evolution of the perturbations.
We shall make this point more explicit below.

\subsection{The Gelaton Model}

In single-field inflationary models with speed of sound $c_s$, {e.g.} in models of DBI inflation,
the equation of motion for the perturbations
\beq
\label{eomgel4}
u''+(c_s^2k^2-\frac{2}{\tau^2})\, u = 0
\eeq
has a solution which reads:
\beq
\label{expa3}
u_\mathrm{eff}(k\tau) = Ae^{-\imath c_sk\tau}\left(1-\frac{\imath}{c_sk\tau}\right) = -\imath A\left[\frac{1}{c_sk\tau}
+\frac{1}{2}c_sk\tau-\frac{\imath}{3}c_s^2(k\tau)^2 -\frac{1}{8} c_s^3(k\tau)^3+\ldots\right] \, .
\eeq

\label{recap}

The authors of Ref.\ \cite{Tolley:2009fg} use the hierarchy $3\eta_{ss}-2\xi^2\gg1$,
and the resulting hierarchy of the eigenvalues of the matrix $Q$ in (\ref{qdef})
to divide each component
$u_\sigma$ and $u_s$ of the solution $\vec u$ of (\ref{eomgel0}) into a heavy and a
light mode.
After a time given by $-k\tau=m/H$, where $m$ is the mass of the heavy mode,  this mode decays as $a^{-1/2}$, oscillating
rapidly.
For the light mode one gets, neglecting the double time derivative in the lower component of (\ref{eomgel0}):
\beq
\label{eq:decoup1}
\left(k^2+\frac{1}{\tau^2}(3\eta_{ss}-2\xi^2-2)\right)u_s^\mathrm{light} = \frac{2\xi}{\tau}\, \partial_\tau u_\sigma^\mathrm{light} +
\frac{2\xi}{\tau^2} \, u_\sigma^\mathrm{light} \, .
\eeq
Neglecting $k^2$ and plugging the solution into the upper component of (\ref{eomgel0}), one finally arrives at
\beq
\label{ulightfinal}
\left(1+\frac{4\xi^2}{3\eta_{ss}-2\xi^2-2}\right)\left( \frac{\mathrm{d}^2}{\mathrm{d}\tau^2} -\frac{2}{\tau^2} \right)
u_\sigma^\mathrm{light} + k^2 u_\sigma^\mathrm{light} = 0 \, .
\eeq
The important point to notice is that this is exactly of the form (\ref{eomgel4}), with\footnote{In the last step, we omitted $-2$ in the denominator, which is justified for $\xi\gg1$ unless $m_\mathrm{gel}^2/H^2\ll 1$.}
\beq
\label{eq:cs1}
c_s^2 =
1 - \frac{4\xi^2}{3\eta_{ss}+2\xi^2-2} \approx 1- \frac{4\xi^2 }{m_\mathrm{gel}^2/H^2 + 4\xi^2} \, ,
\eeq
where,
following the notation of \cite{Tolley:2009fg}, we denote the mass of the heavy mode as
$m_\mathrm{gel}^2 \equiv C_{ss}\approx (3\eta_{ss}-2\xi^2)H^2$, and we also note that $\xi^2 = e^{2b} \, b_\phi^2 \dot{\chi}^2$.
Thus, we can identify $u_\sigma^\mathrm{light}$ with $u_\mathrm{eff}$ above.
We also note that the requirement $m_\mathrm{gel}/H\gg 1$ can easily be translated into
a lower bound\footnote{We should mention that in Section \ref{sectionIVB} we will obtain a much more stringent lower bound
on $c_s$.} on the sound speed if we require that $c_s^2$ deviates significantly from 1:
\beq
\label{csbound2}
c_s^2 \gg\frac{1}{\xi^2} \, .
\eeq

By looking at the expression for the gelaton mass, it is easy to see that in order to get a very small sound speed $c_s \ll 1$
one needs $3\eta_{ss} \gtrsim 2\xi^2 \gg 0$ \footnote{More precisely, to get $c_s^2 \sim \delta$, with $\delta$ small,
we must take $3\eta_{ss}\sim 2\xi^2 (1+2\delta)$. Equivalently, in terms of the gelaton mass, $2\xi \sim \frac{1}{\sqrt\delta}\frac{m_\mathrm{gel}}{H}.$}.
To make the analysis more transparent, we can rewrite the sound speed
in terms of the parameter
$\nu = 3\eta_{ss}/\xi^2-1$
we introduced to keep track of the mass hierarchy in the mass effective matrix:
\beq
\label{csnu}
c_s^2 = 1-\frac{4}{\nu+3} \, .
\eeq
Clearly $\nu \gg 1$ gives the standard $c_s=1$ result, while in order to obtain $c_s \ll 1$ one needs $\nu \sim 1$,
which corresponds to no hierarchy of masses in the effective mass matrix $\mathcal{M}$.

It would be instructive to compare $u_\mathrm{eff}$, defined in (\ref{expa3}), to the full
solution of the equations of motion of the original two-field model (\ref{eomgel0}).
The latter solution, in the form of a power series in $(k\tau)$,
is presented in detail in the Appendix. That expansion is consistent
with (\ref{expa3}) for a large number of terms. Also, our numerical
results in Section \ref{sec:subnum} will show that (\ref{expa3}) is an excellent approximation
to the full solution at sufficiently late times.
However, we would like to stress that although the parameter $c_s$ in eq.\ (\ref{eomgel4}) has the interpretation of a sound speed,
and enters the dispersion relation accordingly, its presence does {not} guarantee the existence of the corresponding
sound horizon at $-k\tau\sim 1/c_s$: in principle, the perturbations can cross
this scale before the effective single-field theory with a nontrivial sound speed is applicable.

\label{sec:subsfcs}

Before moving on, it is interesting to compare the predictions of the gelaton model for the
normalization of the curvature perturbations with those coming from DBI inflation,
collected in Table \ref{tgel}.
In the effective theory of the gelaton obtained in the manner shown above, the canonically normalized field
corresponding to the curvature perturbations is
$u_\mathrm{eff}=a\,Q_\sigma/c_s$.
In contrast, in the case of DBI inflation we have $u_\mathrm{DBI}=a\,Q_\sigma/c_s^{3/2}$
\cite{Langlois:2008wt,Langlois:2008qf,Langlois:2009ej,RenauxPetel:2009sj}.
Both functions $u_\mathrm{eff}$ and $u_\mathrm{DBI}$
have the form (\ref{expa3}) with $A=1/\sqrt{2c_sk}$, as required by the Wronskian condition.
We therefore conclude that the predictions for the normalization of the power spectrum  $\mathcal{P}_{Q_\sigma}$
in the gelaton model are larger by a factor of $1/c_s$ than those of DBI inflation with the same sound speed $c_s$.
In fact, this introduces an explicit dependence of the power spectrum $\mathcal{P}_{Q_\sigma}$ of the $Q_\sigma$ fluctuations on $c_s$.
In contrast, in models of DBI inflation the various factors of $c_s$ cancel
and one recovers the single-field result, as shown in Table \ref{tgel}.
The `missing' factor of $1/c_s$ is, however, recovered upon expressing this result in terms of the
power spectrum $\mathcal{P}_\mathcal{R}$ of the curvature perturbations, as in  DBI inflation the definition of the slow-roll parameter $\epsilon$
involves $c_s $ \cite{Langlois:2008wt,Langlois:2008qf,Langlois:2009ej,RenauxPetel:2009sj}.
We see that, despite the formal similarity of the final results,
the intermediate steps leading to them are different. These differences can be traced back
to the factors of $c_s^{1/2}$ and $c_s^{3/2}$ multiplying the background inflaton field  and
its fluctuation in the expansion of the DBI action \cite{Langlois:2008wt,Langlois:2008qf,Langlois:2009ej,RenauxPetel:2009sj},
while in the action of the gelaton model these factors
are equal to 1 and $c_s$, respectively. Moreover, in DBI inflation the power spectrum
of the isocurvature perturbations is enhanced by a factor of $1/c_s$ with respect to the
curvature perturbations, while there is so such enhancement in the setup considered here.

%%%%

\subsection{From the Two-Field Dynamics to the Single-Field Description}
\label{sectionIVB}

\begin{table}
\begin{center}
\begin{tabular}{|c|c|c|}
\hline
& $\phantom{aa}\mathcal{P}_{Q_\sigma}\phantom{aa}$ & $\phantom{aa}\mathcal{P}_\mathcal{R}=\left(\frac{H}{\dot\sigma}\right)^2\mathcal{P}_{Q_\sigma}\phantom{aa}$ \\
\hline
gelaton model & $\frac{H^2}{4\pi^2c_s}$ & $\frac{H^4}{4\pi^2c_s\dot\sigma^2}=\frac{H^2}{8\pi^2M_P^2c_s\epsilon}$ \\
DBI inflation & $\frac{H^2}{4\pi^2}$ & $\frac{H^4}{4\pi^2\dot\sigma^2}=\frac{H^2}{8\pi^2M_P^2c_s\epsilon}$ \\
\hline
\end{tabular}
\end{center}
\caption{Comparison of the predictions for the power spectra of the gelaton model and models of DBI inflation.
We denote the speed of the inflaton with respect to the cosmic time by $\dot\sigma$. \label{tgel}}
\end{table}

In the course of our discussion, we have identified four physical scales relevant
for the evolution of the cosmological perturbations in our setup, corresponding to:
\beq
-k\tau = \{ \xi, \; \text{max}\left( \frac{m_\mathrm{gel}}{H},\sqrt{\xi}\right), \; \frac{1}{c_s}, \; 1\}.
\eeq
The second scale -- which is nothing but  (\ref{EFTbegins}) rewritten in terms of the gelaton mass --
plays a crucial role in the behavior of the system.
In fact, as we argued in Section \ref{Predictions},
the time
\beq
\label{EFTtime}
-k\tau \sim \mathrm{max}\left(\frac{m_\mathrm{gel}}{H},\sqrt{\xi}\right)
\eeq
signals the {beginning of the regime of validity of the single-field effective field theory} --
earlier, the two modes evolve as if they had equal masses.
For an easy reference, in Table \ref{tab:scales} we summarize briefly the role played by each scale.
Since the most phenomenologically interesting case is that with $c_s \ll 1$, we will analyze the behavior of the perturbations
under this assumption, and consider the effects of the possible relative hierarchies between
the remaining physical scales. We start with simple analytical estimates and continue with numerical analysis supporting our naive estimates.
\begin{table}
\begin{center}
\begin{tabular}{|c|p{12cm}|}
\hline
$-k\tau=$ & characteristics/description \\
\hline
$\xi$ & Both perturbations start decaying as massive modes, if $\xi>m_\mathrm{gel}/H$. \\
$\mathrm{max}\{m_\mathrm{gel},\sqrt{\xi}\}$ &
The curvature perturbation enters the effective single-field theory regime;
the isocurvature perturbation starts/continues decaying as a massive mode. \\
$1/c_s$ & Sound horizon crossing, if the effective single-field theory is applicable. \\
1 & Hubble radius crossing.\\
\hline
\end{tabular}
\end{center}
\caption{A summary of characteristic scales governing the dynamics of the curvature
and the isocurvature perturbations. \label{tab:scales}}
\end{table}

\subsubsection{Simple Analytic Estimates}
\label{sec:subest}

As we saw in \ref{recap}, the requirement of a small sound speed $c_s\ll 1$
is equivalent to the condition $2\xi \gg m_\mathrm{gel}/H$.
We also want to look at models with $m_\mathrm{gel}/H\gg 1$, which
yields an obvious hierarchy
\beq
\label{hier}
\xi \gg \mathrm{max}\left(\frac{m_\mathrm{gel}}{H},\sqrt{\xi}\right) \, .
\eeq
We note that for $\nu\sim 1$  this condition is the same as the requirement $\mu_+ \gg \mu_-$ we used to solve (\ref{btr3m}) perturbatively.
Recall that having $\mu_+ \gg \mu_-$ allowed us to reduce the full perturbation equations to those for two massive modes,
with nearly equal masses $\sim \mu_+ \sim \xi$. Thus, the hierarchy (\ref{hier}) implies that at $-k\tau=\xi$ the evolution of the perturbations
changes qualitatively: instead of decaying as $1/a$ as at very early times,
the wavefunctions start decaying as those of massive modes, i.e.\ as $(1/a)^{3/2}$.
This behavior then continues until $-k\tau = \mathrm{max}\left(\frac{m_\mathrm{gel}}{H},\sqrt{\xi}\right)$.
In the following we will discuss the consequences of the two
possible hierarchies in (\ref{hier}) -- depending on the relative size of $m_\mathrm{gel}/H$ and $\sqrt\xi$ --
for the predictions of the effective single-field theory.
We leave other choices for Section \ref{CAT}, where we will
explore the full range of possibilities.

The time $-k\tau =1/c_s$ signals the sound horizon crossing if the effective single-field theory described by (\ref{eomgel4}) is applicable.
Thus, we expect a different behavior for the system depending on whether a mode crosses  $-k\tau =1/c_s$ {before} or {after} crossing
the time $-k\tau = \mathrm{max}\left(\frac{m_\mathrm{gel}}{H},\sqrt{\xi}\right)$ at which the effective field theory kicks in.
Below we identify the parameter choices corresponding to these two situations, and obtain the resulting
normalizations of the spectra of the curvature perturbations.
\begin{itemize}
\item
We start by considering the case in which the mode crosses $-k\tau = \mathrm{max}\left(\frac{m_\mathrm{gel}}{H},\sqrt{\xi}\right)$
before it crosses $-k\tau=1/c_s$.
It is easy to check that this requirement
implies
that $m_\mathrm{gel}/H \gtrsim \sqrt{\xi}  \gtrsim 1/c_s$,
so the second largest physical scale we identified is simply:
\beq
\mathrm{max}\left(\frac{m_\mathrm{gel}}{H},\sqrt{\xi}\right) \sim \frac{m_\mathrm{gel}}{H} \,.
\eeq
Hence, for $m_\mathrm{gel}/H>\kappa\sqrt{\xi}$, where $\kappa\sim\mathcal{O}(1)$,
the curvature perturbation stops decaying as a massive mode at $-k\tau=m_\mathrm{gel}/H$, entering
the period of the evolution characterized by the effective theory (the gelaton model) discussed above.
The mode is well described by (\ref{eomgel4}), and therefore freezes in at $-k\tau=1/c_s$,
the sound horizon crossing.
\item
If instead the mode crosses $-k\tau = 1/c_s$ first,
the field dynamics is even simpler. Repeating the simple estimates above, we now find that
\beq
\label{case2}
\mathrm{max}\left(\frac{m_\mathrm{gel}}{H},\sqrt{\xi}\right) \sim \sqrt{\xi} \,.
\eeq
More precisely, this case corresponds to the following hierarchy: $m_\mathrm{gel}/H < \sqrt{\xi} <1/c_s$.
Thus, the curvature perturbation stops decaying as a massive mode at $-k\tau\sim\sqrt{\xi}$ and
enters the regime of the effective single-field theory there.
However, since it is already outside the would-be sound horizon, it freezes in immediately.
In this case -- although the effective single-field propagates with non-trivial sound speed $c_s$
and is still well described by (\ref{eomgel4}) -- there is no sound horizon crossing.
\end{itemize}

We are now ready to discuss the effects of the dynamics outlined above on the power spectra, and in particular
of the relative hierarchy between $m_\mathrm{gel}/H$ and $\sqrt\xi$.
First, we note that a period of decay as a massive mode, between times $\tau_1$ and $\tau_2$, suppresses
the wave function by a factor $(a(\tau_1)/a(\tau_2))^{1/2}=(\tau_2/\tau_1)^{1/2}$.
This in turn suppresses the power spectrum by $\tau_2/\tau_1$.
Second, a `premature' freeze-in, occurring at some time $-k\tau=r$ instead of at $-k\tau=1$,
leads to an enhancement of the spectrum by $r^2$.
Hence, using the relations (\ref{eq:cs1}) between $\xi$, $m_\mathrm{gel}$ and $c_s$, we arrive at the following estimate:
\beq
\label{sest}
\frac{\mathcal{P}_\mathcal{R}}{\mathcal{P}_\mathrm{sf}} =
\frac{1}{c_s} \qquad \textrm{for}\,\, m_\mathrm{gel}/H \simgt\sqrt{\xi}
\eeq
and
\beq
\label{sest1}
\frac{\mathcal{P}_\mathcal{R}}{\mathcal{P}_\mathrm{sf}} \sim  \sqrt{\xi} \qquad \textrm{for}\,\, m_\mathrm{gel}/H\simlt \sqrt{\xi} \, .
\eeq
Results (\ref{sest}) and (\ref{sest1}) are among the main findings of our paper.
A look at Section \ref{sec:subsfcs} (e.g. Table I) shows that (\ref{sest}) corresponds to the normalization of the power
spectrum for a single-field inflationary model with a non-trivial sound speed $c_s$.
This comes as no surprise -- the evolution is dictated by (\ref{eomgel4}), with the sound horizon crossing at $-k\tau=1/c_s$.
On the other hand, (\ref{sest1}) corresponds to the case where the effective single-field
theory applies only \emph{after} the curvature perturbation already crossed $-k\tau=1/c_s$ and froze in
-- the relevant physical scale is set by the coupling and there is no proper sound horizon crossing.

Furthermore, the requirement that the sound horizon crossing occurs within the regime of applicability
of the effective theory, $m_\mathrm{gel}/H \simgt\sqrt{\xi}$, leads to a lower bound on the sound speed:
\beq
\label{csbound}
c_s^2 \simgt \frac{1}{4\xi} \, .
\eeq
We would like to stress here that this bound has a different origin
(and it is parametrically stronger) than the bound $c_s^2\gg1/4\xi^2$ in (\ref{csbound2}).
Recalling (\ref{csnu}), the bound (\ref{csbound}) translates into:
\beq
\nu \simgt 1+ \frac{1}{\xi^2} \, .
\eeq
Thus, it is tied to the fact that the hierarchy of the eigenvalues of
the effective mass matrix (\ref{tq2orig}) is lost once we get too close to $\nu=1$.
Although one can push the sound speed closer to zero by increasing $\xi$, we see that there is tension between
lowering $c_s$ significantly and making the strength of the coupling unnaturally high.
Finally, we note
that the agreement of $\mathcal{P}_\mathcal{R}/\mathcal{P}_\mathrm{sf}$ between the full two-field theory with $m_\mathrm{gel}/H \simgt\sqrt{\xi}$
and the effective single-field model of \cite{Tolley:2009fg}
results from a different evolution inside the sound horizon.

\subsubsection{Numerical Analysis}
\label{sec:subnum}

We validate our analytical estimates in
Section \ref{sec:subest} by solving numerically the full set of equations of motion for the
background, as well as for the perturbations; these rather lengthy equations are displayed
e.g.\ in \cite{Lalak:2007vi}.
We shall often plot the results for the evolution of the instantaneous curvature and isocurvature perturbations
expressed in terms of `instantaneous power spectra', defined by
$\mathcal{P}_\mathcal{R}(k)\, \delta(\mathbf{k}-\mathbf{k}') \equiv \frac{k^3}{2\pi^2} \langle\mathcal{R}^\ast_{\mathbf{k}'}
\mathcal{R}_\mathbf{k}\rangle$ and analogously for $\mathcal{S}$, where
the linear perturbations are treated as Gaussian random variables.
We have also verified that the approximation advocated in Section \ref{Ingredients} -- that there
is no implicit time dependence in the parameters entering the equations of motion for the perturbations --
is very good, and that by solving the full equations of motion for the linear perturbations
for $Q_\sigma$ and $\delta s$ \cite{Lalak:2007vi} one obtains results in agreement with
those following from solving the simplified equations
of motion (\ref{eomgel0}) with constant $3\eta_{ss}$ and $\xi$. Since the latter
approach is much faster and convenient,
we employed it in our numerical analysis.
We would also like to recall that since we are very close to the de Sitter solution,
we can use the approximate relation $N=-\log(-1/k\tau)$ to interpret the results.

We present numerical results for the two cases considered in Section \ref{sec:subest}: with
the ``sound horizon crossing'' within/outside the regime of validity of the effective single-field
theory. To facilitate interpretation we chose a rather extreme value of the coupling, $\xi=300$.
The remaining parameters are collected in Table \ref{tab:par}.

\begin{table}
\begin{center}
\begin{tabular}{|c|c|c|c|c|}
\hline
\multicolumn{2}{|c|}{parameters} & \multicolumn{3}{c|}{properties} \\
\hline
$\xi$ & $m_\mathrm{gel}^2/H^2$ & $c_s$ & hierarchy & sound horizon crossing \\
& & & & within single field EFT? \\
\hline
300 & 100 & $\phantom{a}0.016\phantom{a}$ & $m_\mathrm{gel}/H < \sqrt{\xi} <1/c_s$ & NO \\
$\phantom{a}300\phantom{a}$ & 5000 & 0.12 & $m_\mathrm{gel}/H > \sqrt{\xi} >1/c_s$ & YES \\
\hline
\end{tabular}
\end{center}
\caption{Parameters of the models used in the numerical analysis. \label{tab:par}}
\end{table}

\begin{figure}
\begin{center}
\includegraphics*[height=7cm]{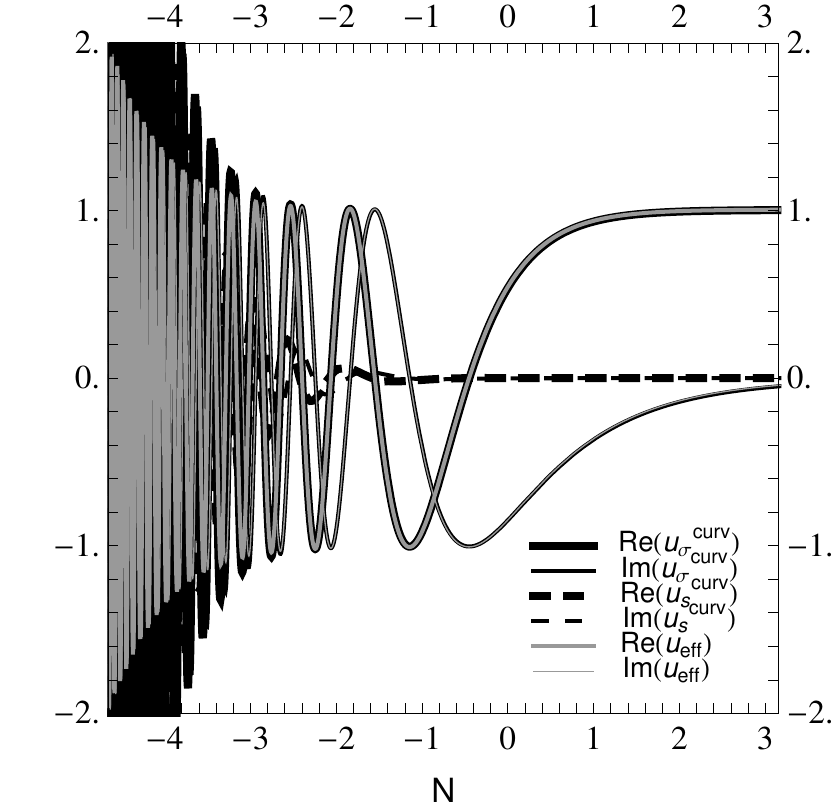}
\hspace{0.3cm}
\includegraphics*[height=7cm]{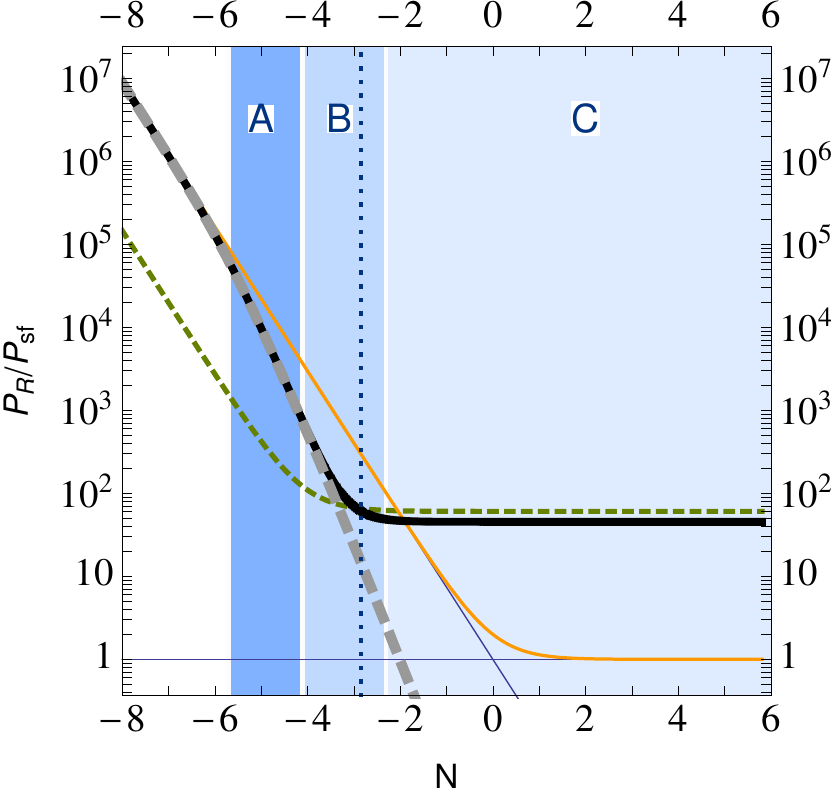}
\end{center}
\caption{Numerical results for the evolution of perturbations calculated in the full two-field system
(from (\ref{eomgel0}) in the limit of constant $\xi$ and $3\eta_{ss}$)
for the first set of parameters in Table \ref{tab:par}:
$\xi=300$ and $m^2_\mathrm{gel}/H^2=100$. For this case $c_s^2=2.7\cdot10^{-4}$.
\underline{Left panel:}
black lines are the real and imaginary parts of both components of the
solution of eq.\ (\ref{eomgel0}), corresponding to final curvature perturbations. Gray lines show the
real and imaginary parts of the effective single-field solution $u_\mathrm{eff}$ given by (\ref{expa3}).
Normalizations and overall phases are chosen so that the imaginary
parts vanish toward the end of inflation.
\underline{Right panel:}
evolution of the instantaneous curvature and isocurvature perturbations shown in terms of the instantaneous power spectra,
as described in the text. $N=0$ corresponds to the Hubble radius exit.
Shaded areas, labeled A, B and C, indicate the ranges  $-k\tau<\xi$, $-k\tau<1/c_s$ and $-k\tau< m_\mathrm{gel}/H$, respectively,
while the vertical dotted line corresponds to $-k\tau = \sqrt\xi$.
The black solid (gray dashed) lines correspond to curvature (isocurvature) perturbations;
the short-dashed (green) line is the solution (\ref{expa3}) with normalization satisfying the Wronskian condition.
All results are normalized to the final value of the single-field, $c_s=1$ solution, which is shown as
the solid gray (orange) line ending at 1.
\label{fgel2}}
\end{figure}

The results for the first set of parameters are shown in Figure \ref{fgel2}.
For this case we have $m_\mathrm{gel}/H<\sqrt\xi <1/c_s$.
According to our discussion around eq.\ (\ref{case2}), the passage to the effective theory should be made at $-k\tau\sim\sqrt{\xi}$
rather than at $-k\tau\sim m_\mathrm{gel}/H$, since the former value is larger.
The left panel of Figure \ref{fgel2} shows the real and imaginary parts of both components, $u_1$ and $u_2$, of the
final {curvature} perturbations
(black lines in the figure),
as well as the real and imaginary parts (gray in the figure) of the solution $u_\mathrm{eff}$ (\ref{expa3})
corresponding to the single-field model with sound speed (\ref{eq:cs1}).
For $-k\tau<\sqrt{\xi}$ (or equivalently $N \gtrsim -2.8$) the solution $u_\mathrm{eff}$ is indeed a good approximation to $u_\sigma$,
justifying the use of the single-field effective description with non-trivial speed of sound, at sufficiently late times.
On the other hand, the oscillations at early times are due to a completely different leading time dependence
of the solution.  As long as the two-field evolution does not reduce to a single-field theory, the sound speed of each field is always $c_s=1$,
and therefore $u_\sigma$ oscillates as $e^{-\imath k\tau}$, and not as $e^{-\imath c_sk\tau}$.
The right panel shows the results for the instantaneous power spectra of the curvature and
isocurvature perturbations, normalized to the value of the curvature perturbations
in single-field models at the end of inflation (\ref{psf}).
The color-coded areas A, B, C denote, respectively, the conditions
$-k\tau <\xi$, $-k\tau < 1/c_s$ and $-k\tau <m_\mathrm{gel}/H$.
The time $-k\tau \sim \sqrt\xi$ is denoted by the dotted vertical line.
The graph shows clearly that at the beginning of region A both modes start decaying as massive modes, with nearly equal masses, $\mathcal{P}_\mathrm{R},\mathcal{P}_\mathrm{S}\sim e^{-3N}$.
This behavior then stops at $-k\tau \sim \sqrt\xi$.
With this choice of $\xi$ and $m_\mathrm{gel}$, the mode crosses $1/c_s$ {before} it enters the region where $-k\tau >\sqrt{\xi}$,
so there is no sound horizon crossing within the regime of validity of the effective theory.
This does {not} by itself contradict the validity of the effective theory: after
$-k\tau =\sqrt{\xi}$ the wave function of the curvature perturbations quickly assumes the asymptotic
constant value.
Finally, at late times the power spectrum which we obtained (the black solid line in the right panel) is slightly suppressed relative to that
predicted by an effectively single-field model (the short, green dashed line).
This suppression is precisely what is expected from (\ref{sest1}): when
$\sqrt\xi < 1/c_s$ as in this case,  (\ref{sest1}) is smaller than (\ref{sest}).

\begin{figure}
\begin{center}
\includegraphics*[height=7cm]{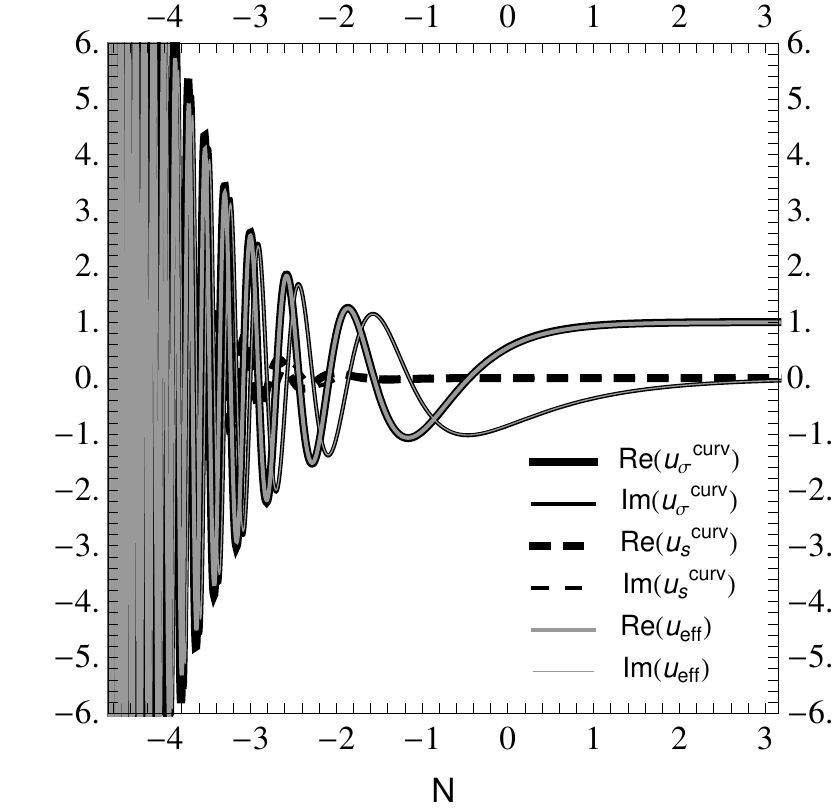}
\hspace{0.3cm}
\includegraphics*[height=7cm]{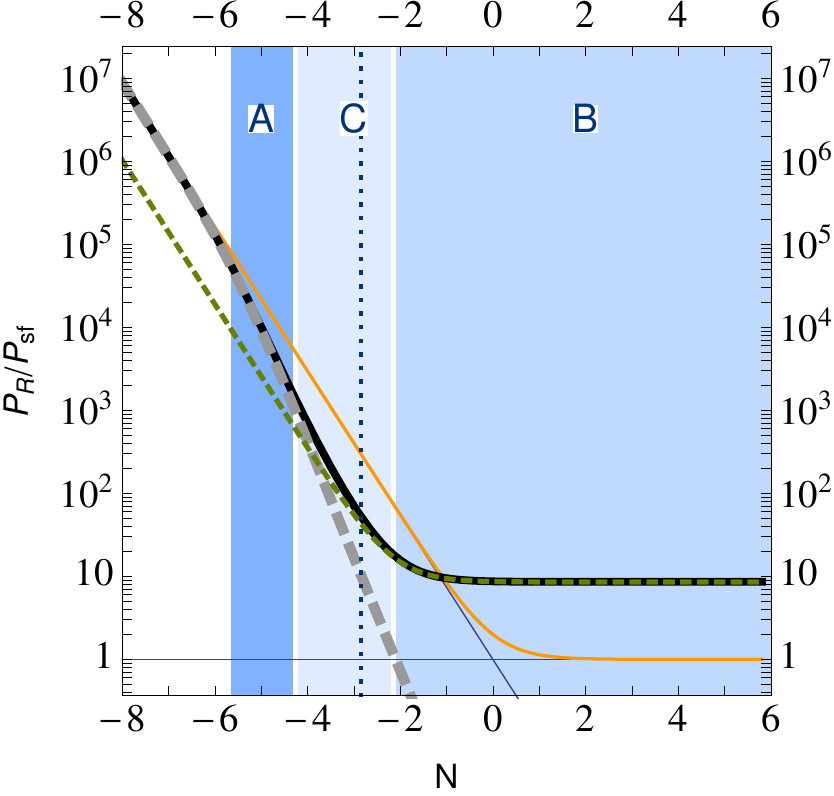}
\end{center}
\caption{The same as in Figure \ref{fgel2}; numerical results for the evolution of perturbations calculated from (\ref{eomgel0}) in the limit
of constant $\xi$ and $3\eta_{ss}$ for the second set of parameters in Table \ref{tab:par}: $\xi=300$
and $m^2_\mathrm{gel}/H^2=5000$ (corresponding to $c_s^2=1.4\cdot10^{-2}$).
\label{fgel2b}}
\end{figure}

Figure \ref{fgel2b} shows analogous results for the second set of parameters in Table \ref{tab:par},
for which $m_\mathrm{gel}/H>\sqrt\xi >1/c_s$.
Notice that with this choice of parameters the sound speed is $c_s \sim 0.12$, about a factor of ten larger than in Figure \ref{fgel2}.
Region A still shows clearly the beginning of the period of decay of the perturbations as massive modes, with nearly equal masses.
Also, since $\sqrt{\xi}<m_\mathrm{gel}/H$, the mode now crosses the $1/c_s$ line {after} crossing $-k\tau = m_\mathrm{gel}/H$,
with the latter moment marking the beginning of the validity of the effective theory.
Thus, the sound horizon crossing lies within the regime of applicability of the effective theory,
and the curvature mode freezes in at $-k\tau=1/c_s$. Furthermore, (\ref{sest}) properly characterizes the power spectra, as expected.

\begin{figure}
\begin{center}
\includegraphics*[height=7cm]{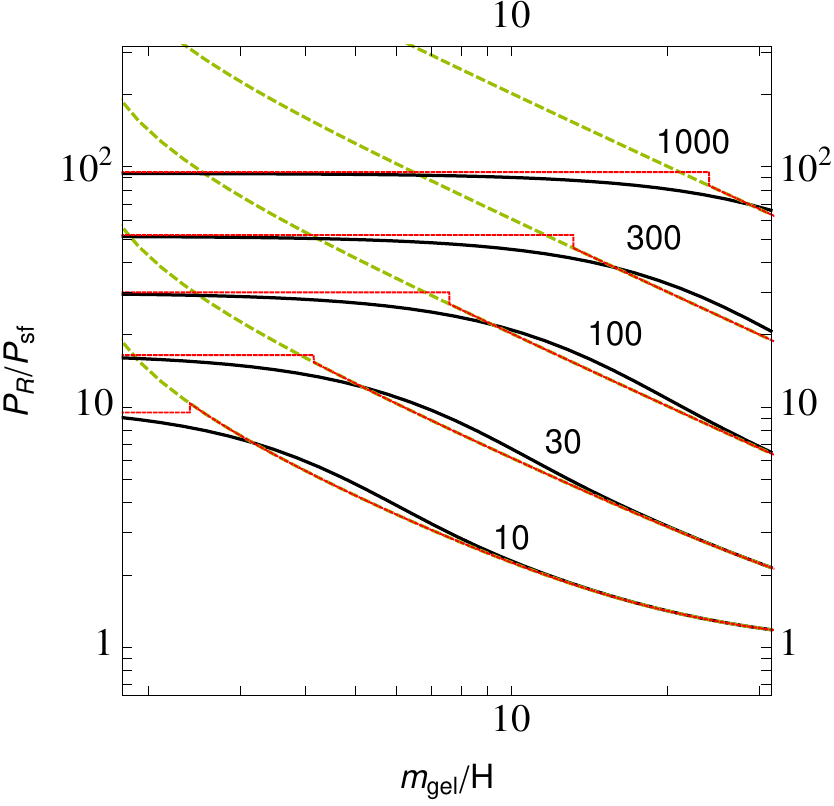}
\end{center}
\caption{Predictions for the power spectrum of the curvature perturbations (normalized
to the single-field result)
as a function of the gelaton mass parameter $m_\mathrm{gel}^2=(3\eta_{ss}-2\xi^2)H^2$ for
different values of $\xi=10,\,30,\,100,\,300,\,1000$.
Black solid lines show the numerical results for the evolution of the perturbations in the full two-field system,
calculated from (\ref{eomgel0}) in the limit of constant $\xi$ and $3\eta_{ss}$.
Red dotted lines correspond to our analytic estimates
(\ref{sest}) and (\ref{sest1}) (with the proportionality constant in (\ref{sest1}) set to 3).
Green short-dashed lines correspond to the estimate (\ref{sest}) outside the limits of its applicability.
\label{fgel3}}
\end{figure}

Figures \ref{fgel2} and \ref{fgel2b} illustrate the claim we have made in Section \ref{sec:subest} -- that
the applicability of the predictions of the effective single-field theory
requires the curvature mode to cross the effective sound horizon only once the effective theory is valid.
When this is not the case, the relevant scale controlling observables such as the power spectrum (\ref{sest1}) is set by
the coupling.
We corroborate this finding by comparing in Figure \ref{fgel3},
for various values of $m_\mathrm{gel}/H$ and $\xi$,
the predictions for the power spectra of the curvature perturbations
calculated numerically from (\ref{eomgel0}) in the limit of constant $\xi$ and $3\eta_{ss}$
(solid black lines) with the estimates
(\ref{sest}) and (\ref{sest1}), shown as
%gray/
red dotted lines (the steps mark the end
of the domain of applicability of one solution and the beginning of another
at $m_\mathrm{gel}/H\sim \sqrt{\xi}$).
The results are different in these two regimes, as we illustrate by extrapolating the predictions
of the effective single-field theory (\ref{sest}) to parameters for which there is
no sound horizon crossing within the regime of validity of the effective theory (dashed green lines).

\subsubsection{Summary of Results}

In summary, we agree with Ref.\ \cite{Tolley:2009fg} that a proper decoupling of a
massive isocurvature mode
will lead to an effective theory describing a single scalar field whose perturbations propagate
with a speed of sound $c_s^2<1$ (see also \cite{Achucarro:2010jv} for a related discussion).
However, the decoupling of the heavy mode occurs when the physical wave numbers cross
the larger of the scales $\sqrt{\xi}H$ and $m_\mathrm{gel}$, so
$m_\mathrm{gel}$ alone does not
give an unambiguous characterization of the decoupling.
We also note that the period of the evolution described by a single-field effective theory with
a non-trivial speed of sound is preceded by a stage during which the
two coupled curvature and isocurvature perturbations behave as massive modes.
This suppresses the power spectra and thereby allows for obtaining the normalization
of the power spectra predicted by the effective single-field theory
(in particular, models of DBI inflation with the same speed of sound).

%%%%%%%%%%%%%%%%%%%%
%%%%%%%%%%%%%%%%%%%%
%%%%%%%%%%%%%%%%%%%%

\section{Two-Field Dynamics: Catalogue of Possibilities}
\label{CAT}

\begin{figure}
\begin{center}
\includegraphics*[height=7cm]{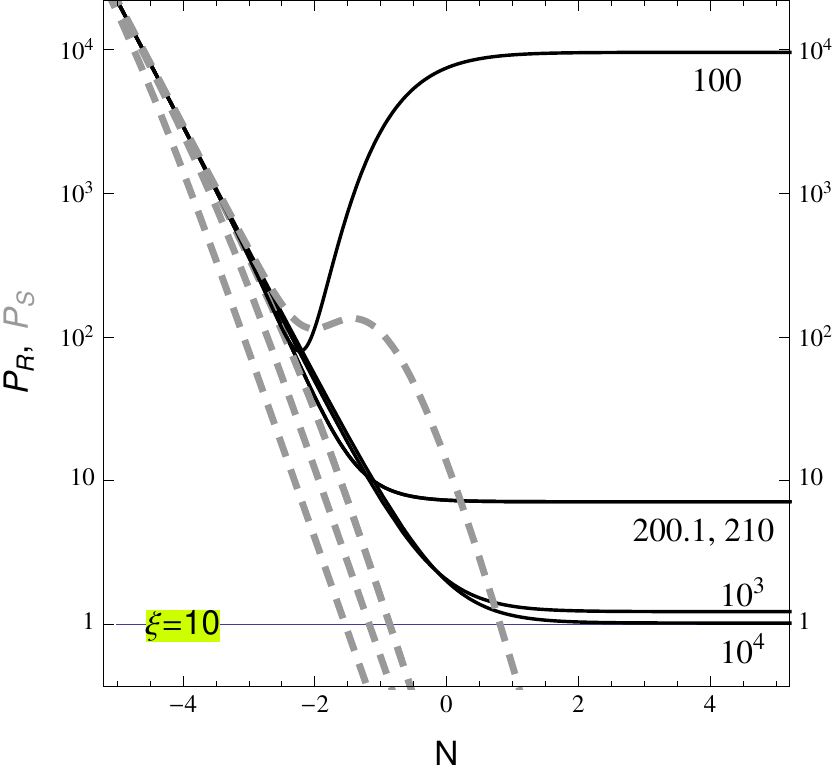}
\hspace{0.5cm}
\includegraphics*[height=7cm]{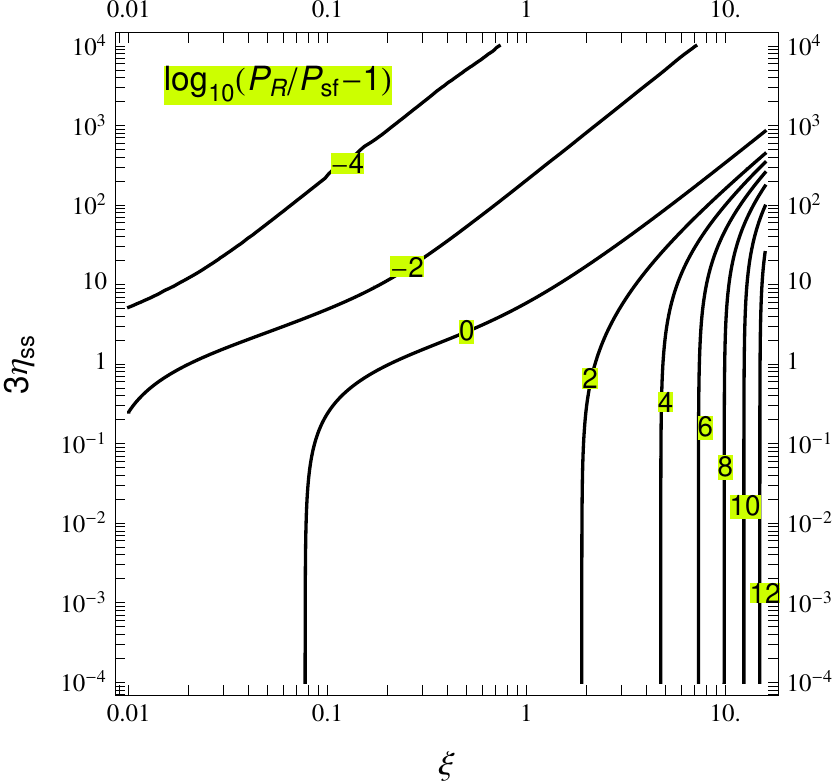}
\end{center}
\caption{\underline{Left panel:} Evolution of the instantaneous curvature and isocurvature perturbations, obtained by
solving (\ref{eomgel0}), shown in terms of the instantaneous power spectra, as described in the text.  Solid (dashed) lines are the (iso)curvature perturbations. We show results for
$\xi=10$ and $3\eta_{ss}=100,\,200.1,\,210,\,10^3,\,10^4$ (top to bottom for the curvature perturbations, left to right for the isocurvature perturbations). The overall normalization is such that 1 corresponds to
a late-time power spectrum for a massless scalar field in the de Sitter space.
\underline{Right panel:} Predictions for the curvature perturbations normalized to the single-field result
given in (\ref{psf}), obtained by
solving (\ref{eomgel0}).
In both panels $N=0$ corresponds to the Hubble radius exit, the results are obtained in the limit of
constant $\xi$ and $\eta_{ss}$.
 \label{fcat}  \label{falt}}
\end{figure}

Depending on the hierarchy between the parameters entering the effective `mass matrix' $\mathcal{M}$ in (\ref{tq2orig}),
we can distinguish different patterns of the evolution of the perturbations in the strong coupling
regime $\xi^2\gg1$.
For concreteness, we choose $\xi=10$.
The results are shown in Figure \ref{fcat} in terms of the potential parameter $\eta_{ss}$ rather
than the effective mass parameter $m_\mathrm{gel}$, which we used previously.

The simplest situation corresponds to $\eta_{ss}\gg \xi^2\gg 1$,  \emph{i.e.} $\nu \gg 1$.
In this case the isocurvature perturbations are very heavy,  and they decay even before
Hubble radius crossing, at $k\tau\sim-\sqrt{3\eta_{ss}}$.
The presence of a coupling to the curvature perturbations, which is strong but still
negligible compared to the mass scale, does not change qualitatively the usual picture.
The curvature perturbations freeze in very close to the Hubble radius crossing, $k\tau\sim-(1+\frac{4\xi^2}{3\eta_{ss}})$,
in accord with the single-field solution of \cite{Tolley:2009fg}, and their amplitude is enhanced
by a factor of $\frac{8\xi^2}{3\eta_{ss}}$ \cite{Achucarro:2010jv}. In Figure \ref{fcat}, this possibility
corresponds to lines with
$3\eta_{ss}=10^4,\,10^3$.

There is also the possibility that $3\eta_{ss}\simgt 3\eta_{ss}-2\xi^2\gg 1$,
which corresponds to the
original `small sound speed' regime of the
gelaton model of Ref.\ \cite{Tolley:2009fg}.
Since we have devoted Section \ref{GEL} to a detailed discussion of this case,
we do not repeat the conclusions here.

 Decreasing the gelaton mass squared $(3\eta_{ss}-2\xi^2)H^2$ to negative values
 leads to $c_s^2<0$. However,
this effective speed of sound is not a fundamental parameter of the theory and,
in particular, $c_s^2<0$ does not indicate any troublesome instability.
This illustrated in Figure \ref{fcat},
where the predictions
for $3\eta_{ss}=200.1$ and $3\eta_{ss}=210$ overlap,
while $c_s^2=-5\cdot 10^{-3}$, $C_{ss}=0.1H^2$ ($c_s^2=2\cdot 10^{-2}$, $C_{ss}=10H^2$) for the
former (latter) parameter choice.

Finally, one can consider
$2\xi^2> |\eta_{ss}|$. Then the  effective `mass matrix' $\mathcal{M}$ in (\ref{tq2orig})
has two large eigenvalues of opposite signs, similar in magnitude.
Because of this large negative mass squared, much larger than the gravitational one,
at $-k\tau=|\xi|$ the evolving modes encounter an instability.
The evolution of the perturbations in this case is shown by the $3\eta_{ss}=100$ line in Figure \ref{fcat}.
This possibility may be interesting for more realistic inflationary model building, as it allows
to circumvent the common gravitino mass problem
\cite{Kallosh:2004yh},
typical of inflationary models embedded in supergravity.
If the slow-roll parameter $\epsilon\sim\mathcal{O}(10^{-2})$, the
observed normalization of the curvature perturbations points to the scale of the potential
$V\sim10^{14}\,\mathrm{GeV}$, which is much larger than what is needed for
softly broken TeV-scale supersymmetry.
While
it is possible to
reduce the scale of the potential by making the potential extremely flat, such models are
very fine-tuned. Here, we can increase the amplitude of the curvature perturbations
by many orders of magnitude above $\mathcal{P}_\mathrm{sf}$ (corresponding to 1 in Figure \ref{fcat}),
thereby decoupling it from $V$ and $\epsilon$.
We illustrate this point in Figure \ref{falt}, which shows the contours of the predicted
$\mathcal{P}_\mathcal{R}$, normalized to $\mathcal{P}_\mathrm{sf}$, on the $(\xi,3\eta_{ss})$ plane
of parameters. We see that allowing for $\xi\sim10$, one can obtain a spectrum of
curvature perturbations enhanced by 10 orders of magnitude compared to the single-field case.

%%%%%%%
%%%%%%%
\section{Discussion and Conclusions}
\label{Discussion}
%%%%%%%
%%%%%%%

In summary, we have studied an extremely simple
model of two-field inflation with non-canonical kinetic terms, described by the Lagrangian (\ref{ourL}).
Rather than exploring the rich phenomenology arising from different inflationary trajectories, we have focused on a special
trajectory, for
which the evolution of the perturbations is particularly simple.
This choice also allowed us to study the case of a large coupling $\xi$ between the curvature and isocurvature perturbations.
We were particularly interested in whether -- at strong coupling -- the analysis of the full two-field system
would lead to effects that might not be captured by the approach of \cite{Tolley:2009fg},
in which a heavy isocurvature mode was integrated out, yielding an effective
description in terms of a single field with non-trivial sound speed $c_s<1$.

To this end, we have solved the equations of motion for the perturbations both inside
and outside the Hubble radius, using a variety of methods (both analytic and numerical), and determined the
power spectra of the perturbations.
We found that
the power spectrum of the curvature perturbations $\mathcal{P}_\mathcal{R}$ is
enhanced with respect to the spectrum $\mathcal{P}_\mathrm{sf}$
of an ordinary single-field model
(with the same scale of the potential and value of the slow-roll parameter $\epsilon$):
$$
\frac{\mathcal{P}_\mathcal{R}}{\mathcal{P}_\mathrm{sf}} \approx \left\{
\begin{array}{ll}
\frac{1}{c_s} & \qquad\textrm{for sound horizon crossing within the effective theory} \\
\sqrt{\xi} & \qquad\textrm{otherwise.}
\end{array}
\right.
$$
We have seen that these predictions stem from two effects: a period of decay of both the curvature
and isocurvature perturbations as modes with masses $\sim\xi H$ (which tends to suppress $\mathcal{P}_\mathcal{R}$), followed by a
freeze-in of the curvature perturbations 
before the Hubble radius crossing (which leads to an enhancement of $\mathcal{P}_\mathcal{R}$).
Whether the sound horizon crossing can be realized within the effective field theory
depends on the hierarchy of the physical scales in the equations of motion of the full two-field theory.

If the sound horizon crossing occurs within the regime of validity of the
effective theory, the predictions for $\mathcal{P}_\mathcal{R}$
coincide with those of DBI inflation models with the same $c_s$, thanks to the two effects mentioned above.
Interestingly, requiring the sound horizon crossing to lie within the effective theory
leads to a lower bound on the sound speed, $c_s^2 > \frac{1}{4\xi}$. Thus,
although $c_s$ can be made very small, this comes at the cost of having to make $\xi$ larger,
and possibly unnaturally large.

We have also encountered situations in which the would-be sound horizon lies outside the
domain of applicability of the effective single-field theory.
Although in this case the spectrum is still enhanced with respect to an ordinary single-field model, the enhancement is now
smaller, and is controlled by the coupling.
We emphasize that -- despite the lack of a `proper' sound horizon
crossing in this case -- the late-time dynamics can still be described
in terms of an effective single field with non-trivial sound speed.
However, the moment at which the passage to the single-field theory occurs is not described by
the parameters of the effective theory, such as the speed of sound or the mass scales, but rather by the coupling.

Finally, we have attempted to survey more broadly inflationary models with a large coupling
between curvature and isocurvature perturbations. We have found regions in the
parameter space in which the perturbations exhibit a momentary tachyonic growth at
the Hubble radius crossing. This effect can raise the amplitude of the curvature perturbations
by orders of magnitude above the single-field estimate based on the scale of the inflationary
potential and the size of the slow-roll parameters (\ref{psf}). This would in turn reduce
the apparent mismatch between the scale of inflation and the scale of supersymmetry breaking
in supergravity models.

We note that our results could not have been anticipated in the general effective models
of multi-field inflation, constructed along the lines of \cite{Senatore:2010wk}. This is because
the analysis of \cite{Senatore:2010wk} invokes symmetries
which constrain the allowed terms in the Lagrangian. In particular,
our non-canonical kinetic term in (\ref{ourL}) does not respect such symmetries.

A large coupling between the curvature and the isocurvature perturbations stems from the
fact that the potential makes the inflationary trajectory non-geodesic in the field space
\cite{Achucarro:2010jv,Achucarro:2010da}. While such a transient feature might also
arise with canonical kinetic terms, through a fast turn of the trajectory, our use of
non-canonical kinetic terms allows such a coupling to persist almost unchanged for many efolds of
inflation, making the scenario analytically tractable.

For linear perturbations, we have been able to perform a resummation of the effects of the
large coupling between the curvature and isocurvature perturbations.
We concede that it would be interesting to extend our analysis beyond the linear level
in order to check whether the large coupling
between weakly correlated curvature and isocurvature pertubations can
bring about large non-gaussianities. Since the currently  available computational techniques
at the non-linear level are perturbative, we wish to defer this question to future work.

%%%%%%%%%%%%%%%%%%%%%%%%%%%%%%%%%%%%%%%%%%%%%%%%%%%%%%%%%%%%%%%%%%%%%%%%%%%%%%%%%%
%\newpage

\section*{Acknowledgements}
We are grateful to S.\ Watson for many useful conversations throughout the development of this paper.
We also would like to thank T.\ Avgoustidis, A.~C.\ Davis and R.~H.~Ribeiro for comments during the final stages of this work.
The work of S.C.\ has been supported by the Cambridge-Mitchell Collaboration in Theoretical
Cosmology, and the Mitchell Family Foundation.
Z.L.~and~K.T.~are partially supported by the EC 6th Framework Programme MRTN-CT-2006-035863 and by the MNiSW grant N N202 091839.
K.T.~also acknowledges support from Foundation for Polish Science through its programme
Homing. K.T.~is grateful to DAMTP, University of Cambridge and to the Mitchell Institute, Texas A\&M University for their hospitality and stimulating atmosphere.

\section*{Appendix: Evolution of Curvature and Isocurvature Perturbations}
\label{App}

In terms of the rescaled Mukhanov-Sasaki variables $u_\sigma=Q_\sigma/a$ and $u_s=\delta s/a$, the equations of motion for
the perturbations take the form:
\beq
\label{eomgel0a}
\left[ \left(\frac{\mathrm{d}^2}{\mathrm{d}\tau^2}+k^2-\frac{2}{\tau^2}\right)
+\left(\begin{array}{cc} 0 & \frac{2\xi}{\tau} \\ -\frac{2\xi}{\tau} & 0 \end{array}\right)
\frac{\mathrm{d}}{\mathrm{d}\tau}+
\left(\begin{array}{cc} 0  & -\frac{4\xi}{\tau^2} \\
-\frac{2\xi}{\tau^2} & \frac{1}{\tau^2}(3\eta_{ss}-2\xi^2) \end{array}\right)\right]
\left(\begin{array}{c}u_1 \\ u_2 \end{array}\right)=0 \, .
\eeq
Switching basis via $\vec u=R\, \vec{\mathcal{U}}$, where
$R$ is a yet unspecified $2\times2$ rotation matrix, they become:
\beq
\label{btr1}
R''\,\vec{\mathcal{U}}+2R\,\vec{\mathcal{U}}'+R\,\vec{\mathcal{U}}'' +\frac{2\xi}{\tau}ER\,\vec{\mathcal{U}}'+\frac{2\xi}{\tau}ER'\,\vec{\mathcal{U}}+\left(k^2-\frac{2}{\tau^2}+\frac{1}{\tau^2}Q\right)\vec{\mathcal{U}} = 0 \, .
\eeq
Here the prime denotes differentiation with respect to $\tau$, and the $E$ and $Q$ matrices read:
\beq
E=\left(\begin{array}{cc}0&1\\-1 & 0\end{array}\right)
\qquad\textrm{and}\qquad
Q=\left(\begin{array}{cc}0&-4\xi\\-2\xi & 3\eta_{ss}-2\xi^2\end{array}\right) \, .
\label{qdef}
\eeq
Finally, adopting
\beq
\label{Rmatrix}
R=\left(\begin{array}{cc}\cos(\xi\log(-k\tau))&-\sin(\xi\log(-k\tau)) \\ \sin(\xi\log(-k\tau))& \cos(\xi\log(-k\tau))\end{array}\right),
\eeq
gives $R'=-(\xi/\tau)ER$ and (\ref{btr1}) takes the much simpler harmonic oscillator form
\beq
\label{btr2a}
\vec{\mathcal{U}}''+\left(k^2-\frac{2}{\tau^2}+\frac{1}{\tau^2} R^T \mathcal{M} R\right)\vec{\mathcal{U}} = 0 \, ,
\eeq
where $\mathcal{M}$ is the effective mass matrix shown explicitly in (\ref{tq2orig}).
We shall now discuss in detail various methods of solving (\ref{eomgel0a}) and (\ref{btr2a})
in different limits. These results are referred to throughout Section \ref{Predictions}.

\noindent
{{\bf Evolution deep inside the Hubble radius}}\\
\noindent
At early times, \emph{i.e.} $\tau\to-\infty$, the last term in (\ref{btr2a}) can be neglected, and
the system reduces to that of two uncoupled harmonic oscillators, with solutions $\mathcal{U}^{(i)}_I\sim\delta_{iI}e^{-\imath k\tau}$.
In terms of the wave functions $\vec{u}$ appearing in (\ref{eomgel0a}), these solutions can be written as:
\beq
\label{sol:dhr}
\vec u^{(1)}\sim\left( \begin{array}{c}\cos\xi z \\ \sin\xi z\end{array}\right)\,e^{-\imath k\tau}
\qquad\textrm{and}\qquad \vec u^{(2)}\sim\left( \begin{array}{c}-\sin\xi z \\ \cos\xi z\end{array}\right)\,e^{-\imath k\tau} \, ,
\eeq
where $z=\log(-k\tau)$.
Each of the modes
$\vec u^{(j)}$
satisfies standard commutation relations,
which translate into the Wronskian condition
\beq
\label{wronskian}
\sum_i\left({u^{(j)}_i}^\ast\partial_\tau u_i^{(j)}-u^{(j)}_i\partial_\tau {u_i^{(j)}}^\ast \right) = -\imath\,
\eeq
for $j=1,2$, given that $||\vec u^{(j)}||=1/\sqrt{2k}$ initially.
Thus, we have two properly normalized and
independent perturbations.

\noindent
{{\bf Evolution on super-Hubble scales}}\\
\noindent
At sufficiently late times, $-k\tau<1$, we can neglect the $k$-dependent terms
in (\ref{eomgel0a}). Assuming $u_i=\tilde A_i (-\tau)^p$, $i=\sigma,s$, we are led to the following algebraic constraint for the amplitudes $\tilde A_i$:
\beq
\label{sh1}
\frac{1}{\tau^2}\left( \begin{array}{cc}p(p-1)-2 & 2p \, \xi-4 \, \xi \\
-2p \, \xi-2 \, \xi & p(p-1)-2-2\xi^2+3\eta_{ss}\end{array}\right)\left( \begin{array}{c}\tilde A_\sigma \\
\tilde A_s \end{array}\right)=0\, .
\eeq
The requirement of a vanishing determinant gives four nonzero solutions:
\bea
\label{solpm1}
p= -1 &\qquad\textrm{and}\qquad& \frac{\tilde A_s}{\tilde A_\sigma}=0\,, \\
p= -2 &\qquad\textrm{and}\qquad& \frac{\tilde A_s}{\tilde A_\sigma}=\frac{6\xi}{3\eta_{ss}-2\xi^2}\,, \\
p=\frac{1}{2}(1+\sqrt{9-4(3\eta_{ss}+2\xi^2)}) &\qquad\textrm{and}\qquad& \frac{\tilde A_s}{\tilde A_\sigma}=\frac{-3+\sqrt{9-4(3\eta_{ss}+2\xi^2)}}{4\xi}\,, \\
p= \frac{1}{2}(1-\sqrt{9-4(3\eta_{ss}+2\xi^2)}) &\qquad\textrm{and}\qquad& \frac{\tilde A_s}{\tilde A_\sigma}=\frac{-3-\sqrt{9-4(3\eta_{ss}+2\xi^2)}}{4\xi} \, .
\eea
Only (\ref{solpm1}), corresponding to $p=-1$, describes a growing mode, \emph{i.e.}\ a mode which freezes in
after Hubble radius crossing. Since for this mode $\tilde A_s=0$, we can associate it with the growing
mode of the curvature perturbation.
The last two solutions are characteristic of a massive mode with an effective mass squared
\beq
\label{meff1}
m_\mathrm{eff}^2=V_{ss}+2\xi^2H^2=(3\eta_{ss}+2\xi^2)H^2 = (\nu+3)\xi^2H^2\,,
\eeq
which we associate with the isocurvature mode.

\noindent
{{\bf Power series solution}}\\
\noindent
Although the two equations (\ref{eomgel0a})
can be combined into a single (fourth-order) differential equation very similar to one obtained from the Bessel equation,
and reducing to it in the sub-Hubble limit, there do not exist, to our knowledge,
closed form expressions for the solutions of (\ref{eomgel0a}).
On the other hand, it is instructive to go beyond the asymptotic behavior
and to understand the behavior of the growing mode ($p=-1$) solution
closer to the Hubble radius, for a more direct comparison with the gelaton scenarion.
To this end, we take the growing mode to have an expansion of the form:
\bea
\label{expa1}
u_\sigma^{(p=-1)}(\tau) &=& \frac{a_{-1}}{k\tau}+\sum_{n=0}^\infty\frac{a_n}{n!} (k\tau)^n\, ,\qquad a_{-1}\neq 0 \, ,\\
\label{expa2}
u_s^{(p=-1)}(\tau) &=& \sum_{n=0}^\infty \frac{b_n}{n!}(k\tau)^{n+n_0}\, ,\qquad b_0\neq 0 \, ,
\eea
where $n_0$ is for now an unspecified parameter.
This form is consistent with the asymptotic behavior of the solution.
It is easy to check that a nontrivial solution\footnote{Note that the choice $n_0=0$ gives $b_0=0$, which contradicts the assumption
in the mode expansion (\ref{expa2}). If $n_0$ is not an integer, then by substituting (\ref{expa2}) into (\ref{eomgel0a}) and
comparing coefficients of powers of $k\tau$, we again arrive at $b_0=0$.} corresponds to $n_0=1$.
The coefficient $a_{-1}$ is of course arbitrary, and $a_0=0$.
Solving the perturbation equations (\ref{eomgel0a}) recursively order by order, we find
\bea
b_0 &=& \frac{4\xi}{3\eta_{ss}-2\xi^2-2} \, a_1 \, ,\qquad
a_1 = \frac{1}{2} \, \frac{3\eta_{ss}-2\xi^2-2}{3\eta_{ss}+2\xi^2-2} \, a_{-1}\, ,  \\
b_1 &=& \frac{3\xi}{3\eta_{ss}-2\xi^2} \, a_2 \, ,\qquad\qquad
b_2= -\frac{2\xi(3\eta_{ss}-2\xi^2)}{(3\eta_{ss}+2\xi^2-2)(3\eta_{ss}+2\xi^2+4)} \, a_{-1} \, ,  \\
a_3 &=& -\frac{3(9\eta_{ss}^2+6\eta_{ss}(1-2\xi^2)+4(\xi^4-3\xi^2-2))}{4(9\eta_{ss}^2+6\eta_{ss}(1+2\xi^2)
+4(\xi^2-1)(\xi^2+2))} \, a_{-1} \, .
\eea
Continuing this exercise, we arrive at a recursion relation between the $a_n$,
\bea
a_{n+4}&=& - \frac{(n+3)(n+4)}{n(n+5)(n+2)} \frac{1}{3\eta_{ss}+2\xi^2+(n+5)(n+2)} \times \nonumber \\
&&\left[ (n+1)(n+2)^2 \, a_n + n(3\eta_{ss}-2\xi^2 + 2(n+2)(n+4)) \, a_{n+2}\right] \, ,
\label{acoeffgen}
\eea
and a relation defining $b_n$ in terms of $a_j$ :
\beq
b_{n+1} = -\frac{1}{2\xi n(n+2)} \left[ n(n+3)a_{n+2}+(n+1)(n+2)a_n\right] \, .
\eeq
Putting all the ingredients together, the first few terms of the curvature perturbation are:
\bea
\label{eqn2full}
u_\sigma^{(p=-1)}(k\tau) &=& \frac{a_{-1}}{k\tau}\left[1+ \frac{1}{2} \frac{3\eta_{ss}-2\xi^2-2}{3\eta_{ss}+2\xi^2-2}
(k\tau)^2+ \frac{1}{2} \frac{a_2}{a_{-1}} (k\tau)^3 - \right. \nonumber \\
&& \left.
 -\frac{(9\eta_{ss}^2+6\eta_{ss}(1-2\xi^2)+4(\xi^4-3\xi^2-2))}{8(9\eta_{ss}^2+6\eta_{ss}(1+2\xi^2)+4(\xi^2-1)(\xi^2+2))}(k\tau)^4
+ \ldots \right] \, .
%+ \frac{1}{2} \frac{a_2}{a_{-1}} (k\tau)^3 + \frac{1}{6} \frac{a_3}{a_{-1}} (k\tau)^4 \Bigr]
\eea
From this expression one can extract the quantity corresponding to the sound speed of the
gelaton scenario.

Finally, we note that $u^{(p=-1)}_\sigma$ appears to depend on two arbitrary constants, $a_{-1}$ and $a_2$,
which uniquely determine the solution $u^{(p=-1)}_s$.
In principle, one should add to these solutions the decaying modes $u^{(p=2)}_\sigma$ and $u^{(p=2)}_s$,
whose expansions start with terms ${\cal O}\left(\left(k\tau\right)^2\right)$.
A calculation analogous to the one above shows that the only nonzero coefficients in the $p=2$ mode expansion
are those of even powers of $(k\tau)$.
Furthermore, these coefficients are the same (up to an overall normalization constant) as those of the even powers
of the $p=-1$ mode.

\noindent
{{\bf Linking the early- and late-time solutions}}\\
\noindent
Notice that in the strong coupling regime $\xi^2 \gg 1$ eq.\ (\ref{btr2a}) describes the evolution of two massive modes
-- with masses squared $\mu_1^2 \sim\xi^2$ and $\mu_2^2 \sim \nu\xi^2\equiv 3\eta_{ss}-\xi^2$ in Hubble units --
and a rapid rotation of the basis vectors, as visible from computing explicitly $R^T \tilde{Q} R$.
This seems to contradict the observation earlier in this section that the late-time solutions correspond to a massless
mode (giving rise to the scale-invariant spectrum of curvature perturbations) and a heavy mode of
mass given by (\ref{meff1}), $m_\mathrm{eff}^2/H^2=(\nu+3)\xi^2$.
This apparent disagreement is a simple consequence of the fact that,
while eq.\ (\ref{btr2a}) is a result of working in the $\vec{\mathcal{U}}$ basis,
 the above analysis of the super-Hubble evolution was done in the $\vec u$ basis for the perturbations,
\emph{i.e.} was based on solving (\ref{eomgel0a}).
Obviously, these two points of view should agree upon taking into account the change of
basis $\vec u=R \, \vec{\mathcal{U}}$. In the following we show that this is indeed the case.
Since the explanation is particularly instructive and technically feasible
for the $\nu\sim 1$ case  -- no hierarchy of masses in the effective mass matrix (\ref{tq2}) --  we shall
adhere to it for the rest of this section.
As we will see in Section \ref{GEL}, the choice $\nu \sim 1$ is the most relevant phenomenologically
-- it corresponds to a very small sound speed.

Before we proceed further, we recall that a massive mode $u_\mathrm{m}$ in de Sitter space, with mass $m\equiv\mu H$,
obeys the following equation of motion:
\beq
\label{sfmass}
\frac{\mathrm{d}^2u_\mathrm{m}}{\mathrm{d}\tau^2}+\left(k^2+\frac{\mu^2-2}{\tau^2}\right)u_\mathrm{m} = 0 \, .
\eeq
For $-k\tau\ll \mu$, eq. (\ref{sfmass})
can be solved approximately by neglecting the $k^2$ term, giving:
\beq
\label{sfmass1}
u_\mathrm{m} \sim (-k\tau)^{\frac{1}{2}\left( 1-\sqrt{9-4\mu^2}\right)} \quad
\longrightarrow \quad
\sqrt{-k\tau}e^{-\imath\mu\log(-k\tau)} \quad \text{for} \quad \mu \gg 1 .
\eeq
In the following, when discussing the solutions to the perturbation equations (\ref{btr2}),
we shall compare them to (\ref{sfmass1}) to read off the corresponding mass parameters.

In order to solve (\ref{btr2}) it turns out to be convenient to define $\mu_\pm^2 \equiv \half (\lambda_2 \pm \lambda_1)$,
with $\lambda_1,\lambda_2$ the eigenvalues of (\ref{tq2}), giving us:
\beq
\mu_+^2=(\nu+1)\frac{\xi^2}{2} \, , \quad \quad \mu_-^2=\frac{\xi^2}{2}\sqrt{(\nu-1)^2+\frac{36}{\xi^2}} \, .
\eeq
Note that for $\nu$ close to 1 we can approximate both expressions, and obtain:
\beq
\label{apprmu}
\mu_+^2 \sim \xi^2 \, , \quad \quad \mu_-^2\sim \mathrm{max}\left( 3\xi,(\nu-1)  \frac{\xi^2}{2} \right) \, .
\eeq
With these definitions, we can isolate the rapidly oscillating terms and rewrite (\ref{btr2}) as:
\beq
\label{btr3}
\vec{\mathcal{U}}^{\, \prime\prime} +\left[ \left( k^2+\frac{\mu_+^2-2}{\tau^2}\right) +\frac{\mu_-^2}{\tau^2} \left(\begin{array}{cc} -
\cos(2\xi \log(-k \tau)) & \sin(2\xi \log(-k \tau)) \\ \sin(2\xi \log(-k \tau)) & \cos(2\xi \log(-k \tau))\end{array}\right)\right]
\vec{\mathcal{U}} =0 \, .
\eeq
Since we are always working under the assumption of strong coupling, $\xi \gg 1$, our parameter choice $\nu \sim 1$ corresponds to $\mu_+\gg \mu_-$.
This hierarchy allows us to solve (\ref{btr3}) perturbatively,
as the effects of the rapid rotation are suppressed by the relative smallness of $\mu_-$ with respect to $\mu_+$.
We proceed by splitting $\vec{\mathcal{U}}=\vec{\mathcal{U}}_{(0)} + \vec{\mathcal{U}}_{(1)}$,
where $\vec{\mathcal{U}}_{(0)}$ is a positive frequency solution of the unperturbed equation (\emph{i.e.} assuming $\mu_-=0$),
\beq
\label{u0eom}
\vec{\mathcal{U}}_{(0)}^{\, \prime\prime} +\left[ \left( k^2+\frac{\mu_+^2-2}{\tau^2}\right)  \right] \vec{\mathcal{U}}_{(0)} =0 \, ,
\eeq
and $\vec{\mathcal{U}}_{(1)}$ is a special solution of:
\beq
\label{btr4}
\vec{\mathcal{U}}_{(1)}^{\, \prime\prime} +\left[ \left( k^2+\frac{\mu_+^2-2}{\tau^2}\right) \right] \vec{\mathcal{U}}_{(1)}+\frac{\mu^2_-}{\tau^2} \left(\begin{array}{cc} -
\cos(2\xi \log(-k \tau)) & \sin(2\xi \log(-k \tau)) \\ \sin(2\xi \log(-k \tau)) & \cos(2\xi \log(-k \tau))\end{array}\right) \vec{\mathcal{U}}_{(0)} =0 \, .
\eeq
The leading order solution $\vec{\mathcal{U}}_{(0)}$ consists of two independently evolving modes of equal masses $\mu_+ H$
which, for $\mu_+\gg 1$, are of the form
\beq
\vec{\mathcal{U}}_{(0)}^{\,(a,b)}\sim \,\sqrt{-k\tau} \,e^{-\imath \mu_+\log(-k\tau)} \vec e_\pm \, ,
\quad
\quad \vec e_\pm=\left( \begin{array}{c} 1\\ \pm\imath \end{array}\right) \,.
\eeq
Plugging these into (\ref{btr4}), we find the correction $\vec{\mathcal{U}}_{(1)}$ to the leading order result:
\beq
\vec{\mathcal{U}}_{(1)}^{\,(a,b)} \sim \sqrt{-k\tau} \,e^{-\imath (\mu_+ \pm 2\xi)\log(-k\tau)} \vec{e}_\mp\, .
\eeq
Thus, the leading solution $\vec{\mathcal{U}}_{(0)}$ represents massive modes with mass parameter $m^2=\mu_+^2H^2$,
while the first-order contribution $\vec{\mathcal{U}}_{(1)}$ is a combination of modes with squared masses $(\mu_+\pm2\xi)^2H^2$.
It is easy to check that continuing the iterative solution does not introduce any further mass parameters.
This is a remarkably simple result: despite a nondiagonal, time-dependent mass matrix in the
$\vec{\mathcal{U}}$ basis, we see that the solution of the equations of motion for the two fields
is a combination of single-field solutions, obtained by replacing $\mu$ in the massive mode solution (\ref{sfmass1}) with
one of the three mass parameters ${\mu_+,\mu_+ + 2\xi,\mu_+ -2\xi}$.

The modes $\vec{\mathcal{U}}=\vec{\mathcal{U}}_{(0)}+\vec{\mathcal{U}}_{(1)}$ above can be related to the late time solutions $\vec u$
via the transformation $\vec{u}=R\, \vec{\mathcal{U}}$.
In particular, since $R \, \vec{e}_\pm=e^{\mp\xi\log(-k\tau)} \vec{e}_\pm$,
the $R$  rotation changes the argument of the exponential, and {shifts the mass parameters} we identified above by $\mp \xi$.
Thus, the end result of the change of basis is the following shift in the masses of the solutions:
the mass parameters $\mu_+$ and $\mu_+ + 2\xi$, corresponding to
$\vec{\mathcal{U}}_{(0)}^{(a)}$ and $\vec{\mathcal{U}}_{(1)}^{(a)}$,
are lowered to $(\mu_+-\xi)$ and $(\mu_+ + \xi)$.
Similarly, the parameters of the second solution
$\vec{\mathcal{U}}_{(0)}^{(b)}$ and $\vec{\mathcal{U}}_{(1)}^{(b)}$,
given by $\mu_+$ and $\mu_+-2\xi$ respectively, get shifted upwards to $(\mu_++\xi)$ and $(\mu_+-\xi)$.
Hence, we see that the $\vec u$ solution consists of two massive modes with mass parameters $(\mu_+\pm\xi)^2H^2$.
Finally, making use of the fact
\footnote{Recall that the hierarchy $\mu_+\gg\mu_-$, which justified
our perturbative approach for solving (\ref{btr3}), occurs for $\nu\sim 1$, implying (up to small corrections) $\mu_+\sim\xi$.}
that $\mu_+\sim\xi$, we see that
the mode in the $\vec u$ basis with $m^2 \sim (\mu_+-\xi)^2 H^2$ is approximately massless,
while the $m^2 \sim (\mu_+ + \xi)^2H^2$ mode has a mass squared $\sim 4\xi^2 H^2$.
As expected, this is in perfect agreement with what we have found on super-Hubble scales.

The perturbative procedure used above is only justified for $\mu_-^2/(k\tau)^2\ll 1$,
when the modes do not `feel' the full effect of the strong time-dependence encoded in $R^T \tilde{Q} R$.
This approximation breaks down when $|k\tau|\sim {\cal O} (\mu_-)$.
Thus, since $\mu_-^2=\mathrm{max}\{3\xi,(\nu-1)\xi^2/2\}$, we expect that the period during which both
fields decay as massive modes (with nearly equal masses) ends as soon as $-k\tau$ reaches the {\em larger} of the values
$\sim\sqrt{\xi}$ or $\sim\sqrt{\nu-1}\xi$.
This marks the onset of the asymptotic super-Hubble behavior and the
beginning of the regime of validity of an effective single-field description.

%%%%%%%%%%%
%%%%%%%%%%%
%%%%%%%%%%%

\noindent
{{\bf The power series solution and the sound speed}} \\
\noindent
We can also address the validity of the effective theory by comparing the
expansion (\ref{expa3}) with the full solution written in terms of the power series
(\ref{expa1}) described by the recurrence relation (\ref{acoeffgen}).
If the dynamics of the full two-field theory is correctly captured by the single-field effective
theory proposed in \cite{Tolley:2009fg}, the two mode functions should agree {within the regime
of validity of the effective theory}. More precisely,
we should compare the odd powers of $(k\tau)$ in the expansion, since they are absent
in the expansion of the $p=2$ decaying
solution\footnote{The even powers instead are present in both the growing and
the decaying modes, in $u^{(p=-1)}_{\sigma,s}$ and in $u^{(p=2)}_{\sigma,s}$.}.
Thus, the coefficients $a_n$ of (\ref{expa1}) should agree with those of (\ref{expa3}) for odd $n$.
In particular, they should have the same numerical coefficients and the same powers of
$c_s$.
By comparing the expansions (\ref{eqn2full}) and (\ref{expa3}) with $c_s^2$ given by (\ref{eq:cs1}),
we see that this is trivially satisfied for terms linear in $(k\tau)$. As for the $(k\tau)^3$ term,
the factor of $(c_s^2)^2$ should be obtained from:
\beq
-\frac{4a_3}{3a_{-1}} = c_s^4\left(1-\frac{1}{2c_s^2\xi^2} +\ldots \right) \, ,
\eeq
where the ellipsis stands for terms of higher order in $c_s^2$ and $1/\xi^2$.
For $c_s\ll1$, the requirement that $m_\mathrm{gel}$ is much larger than the
Hubble scale can be translated
into $4c_s^2\xi^2\gg1$, so we recover the correct coefficient predicted by the
effective theory\footnote{As follows from Section \ref{sec:subest}, for $m_\mathrm{gel}/H\simgt \sqrt{\xi}$ we obtain an even stronger inequality $c_s^2\xi^2\simgt\xi$, but for the present purpose a weaker, but general, bound is more useful.}.
By looking at the recursive relation (\ref{acoeffgen}), we also see that as long as
$n<m_\mathrm{gel}/H$, the leading contributions to $a_{2n-1}/a_{-1}$ is proportional
to $c_s^{2n}$, so (\ref{expa3}) is a good approximation for $u_\sigma$ corresponding
to the curvature mode.

\vspace{0.5cm}

%\newpage


\begin{thebibliography}{99}

\bibitem{Quevedo:2002xw}
  F.~Quevedo,
  ``Lectures on string/brane cosmology,''
  Class.\ Quant.\ Grav.\  {\bf 19} (2002) 5721
  [arXiv:hep-th/0210292].

\bibitem{McAllister:2007bg}
  L.~McAllister and E.~Silverstein,
  ``String Cosmology: A Review,''
  Gen.\ Rel.\ Grav.\  {\bf 40}, 565 (2008)
  [arXiv:0710.2951 [hep-th]].

\bibitem{Baumann:2009ni}
  D.~Baumann and L.~McAllister,
  ``Advances in Inflation in String Theory,''
  Ann.\ Rev.\ Nucl.\ Part.\ Sci.\  {\bf 59} (2009) 67
  [arXiv:0901.0265 [hep-th]].

\bibitem{Coughlan:1984yk}
  G.~D.~Coughlan, R.~Holman, P.~Ramond and G.~G.~Ross,
  ``Supersymmetry And The Entropy Crisis,''
  Phys.\ Lett.\  B {\bf 140} (1984) 44.

  \bibitem{Holman:1984yj}
  R.~Holman, P.~Ramond and G.~G.~Ross,
  ``Supersymmetric Inflationary Cosmology,''
  Phys.\ Lett.\  B {\bf 137} (1984) 343.

\bibitem{Brustein:1992nk}
  R.~Brustein and P.~J.~Steinhardt,
  ``Challenges for superstring cosmology,''
  Phys.\ Lett.\  B {\bf 302} (1993) 196
  [arXiv:hep-th/9212049].

\bibitem{Dine:1985vv}
  M.~Dine, V.~Kaplunovsky, M.~L.~Mangano, C.~Nappi and N.~Seiberg,
  ``Superstring Model Building,''
  Nucl.\ Phys.\  B {\bf 259} (1985) 549.

\bibitem{Giddings:2001yu}
  S.~B.~Giddings, S.~Kachru and J.~Polchinski,
  ``Hierarchies from fluxes in string compactifications,''
  Phys.\ Rev.\  D {\bf 66}, 106006 (2002)
  [arXiv:hep-th/0105097].

\bibitem{Silverstein:2004id}
  E.~Silverstein,
  ``TASI / PiTP / ISS lectures on moduli and microphysics,''
  arXiv:hep-th/0405068.

\bibitem{Douglas:2006es}
  M.~R.~Douglas and S.~Kachru,
  ``Flux compactification,''
  Rev.\ Mod.\ Phys.\  {\bf 79}, 733 (2007)
  [arXiv:hep-th/0610102].

\bibitem{Komatsu:2008hk}
  E.~Komatsu {\it et al.}  [WMAP Collaboration],
  ``Five-Year Wilkinson Microwave Anisotropy Probe (WMAP)
  Observations:Cosmological Interpretation,''
  Astrophys.\ J.\ Suppl.\  {\bf 180} (2009) 330
  [arXiv:0803.0547 [astro-ph]].

\bibitem{mukhanov}
V.~Mukhanov,
``Physical foundations of cosmology,''
{\it  Cambridge, UK: Univ. Pr. (2005) 421 p}.

\bibitem{planck}
{\tt http://www.rssd.esa.int/Planck}

\bibitem{GrootNibbelink:2000vx}
  S.~Groot Nibbelink and B.~J.~W.~van Tent,
  ``Density perturbations arising from multiple field slow-roll inflation,''
  arXiv:hep-ph/0011325.

    \bibitem{GrootNibbelink:2001qt}
  S.~Groot Nibbelink and B.~J.~W.~van Tent,
  ``Scalar perturbations during multiple field slow-roll inflation,''
  Class.\ Quant.\ Grav.\  {\bf 19} (2002) 613
  [arXiv:hep-ph/0107272].

\bibitem{DiMarco:2002eb}
F.~Di Marco, F.~Finelli and R.~Brandenberger,
``Adiabatic and Isocurvature Perturbations for Multifield Generalized Einstein Models,''
Phys.\ Rev.\ D {\bf 67}, 063512 (2003)
[arXiv:astro-ph/0211276];

  \bibitem{vanTent:2003mn}
  B.~J.~W.~van Tent,
  ``Multiple-field inflation and the CMB,''
  Class.\ Quant.\ Grav.\  {\bf 21} (2004) 349
  [arXiv:astro-ph/0307048].

 \bibitem{DiMarco:2005nq}
F.~Di Marco and F.~Finelli,
``Slow-roll inflation for generalized two-field Lagrangians,''
Phys.\ Rev.\ D {\bf 71}, 123502 (2005)
[arXiv:astro-ph/0505198].

\bibitem{Lalak:2007vi}
Z.~Lalak, D.~Langlois, S.~Pokorski and K.~Turzynski,
``Curvature and isocurvature perturbations in two-field inflation,''
JCAP {\bf 0707} (2007) 014
[arXiv:0704.0212 [hep-th]].

\bibitem{Vincent:2008ds}
  A.~C.~Vincent and J.~M.~Cline,
  ``Curvature Spectra and Nongaussianities in the Roulette Inflation Model,''
  JHEP {\bf 0810} (2008) 093
  [arXiv:0809.2982 [astro-ph]].

    \bibitem{Peterson:2010np}
  C.~M.~Peterson and M.~Tegmark,
  ``Testing Two-Field Inflation,''
  arXiv:1005.4056 [astro-ph.CO].

\bibitem{Cremonini:2010sv}
S.~Cremonini, Z.~Lalak and K.~Turzynski,
  ``On Non-Canonical Kinetic Terms and the Tilt of the Power Spectrum,''
  Phys.\ Rev.\  D {\bf 82} (2010) 047301
  [arXiv:1005.4347 [hep-th]].

\bibitem{Chen:2009zp}
  X.~Chen and Y.~Wang,
  ``Quasi-Single Field Inflation and Non-Gaussianities,''
  JCAP {\bf 1004} (2010) 027
  [arXiv:0911.3380 [hep-th]].


\bibitem{Tolley:2009fg}
  A.~J.~Tolley and M.~Wyman,
  ``The Gelaton Scenario: Equilateral non-Gaussianity from multi-field dynamics,''
  Phys.\ Rev.\  D {\bf 81}, 043502 (2010)
  [arXiv:0910.1853 [hep-th]].

\bibitem{msw}
E.~Silverstein and A.~Westphal,
``Monodromy in the CMB: Gravity Waves and String Inflation,''
Phys.\ Rev.\  D {\bf 78} (2008) 106003
[arXiv:0803.3085 [hep-th]];

\bibitem{msw2}
  L.~McAllister, E.~Silverstein and A.~Westphal,
  ``Gravity Waves and Linear Inflation from Axion Monodromy,''
  Phys.\ Rev.\  D {\bf 82} (2010) 046003
  [arXiv:0808.0706 [hep-th]].

\bibitem{fibre}
M.~Cicoli, C.~P.~Burgess and F.~Quevedo,
``Fibre Inflation: Observable Gravity Waves from IIB String Compactifications,''
JCAP {\bf 0903} (2009) 013
[arXiv:0808.0691 [hep-th]].

\bibitem{Gordon:2000hv}
C.~Gordon, D.~Wands, B.~A.~Bassett and R.~Maartens,
``Adiabatic and entropy perturbations from inflation,''
Phys.\ Rev.\ D {\bf 63}, 023506 (2001)
[arXiv:astro-ph/0009131].

\bibitem{Langlois:2008mn}
  D.~Langlois and S.~Renaux-Petel,
  ``Perturbations in generalized multi-field inflation,''
  JCAP {\bf 0804} (2008) 017
  [arXiv:0801.1085 [hep-th]].
  
\bibitem{Gao}
  X.~Gao,
  ``Primordial Non-Gaussianities of General Multiple Field Inflation,''
  JCAP {\bf 0806} (2008) 029
  [arXiv:0804.1055 [astro-ph]].

\bibitem{Byrnes:2006fr}
  C.~T.~Byrnes and D.~Wands,
  ``Curvature and isocurvature perturbations from two-field inflation in a
  slow-roll expansion,''
  Phys.\ Rev.\  D {\bf 74}, 043529 (2006)
  [arXiv:astro-ph/0605679].

\bibitem{Berglund:2009uf}
  P.~Berglund and G.~Ren,
  ``Multi-Field Inflation from String Theory,''
  arXiv:0912.1397 [hep-th].

\bibitem{dbisky}
  M.~Alishahiha, E.~Silverstein and D.~Tong,
  ``DBI in the sky,''
  Phys.\ Rev.\  D {\bf 70} (2004) 123505
  [arXiv:hep-th/0404084].

\bibitem{Burgess:2002ub}
  C.~P.~Burgess, J.~M.~Cline, F.~Lemieux and R.~Holman,
  ``Are inflationary predictions sensitive to very high energy physics?,''
  JHEP {\bf 0302}, 048 (2003)
  [arXiv:hep-th/0210233].

\bibitem{Langlois:2008wt}
  D.~Langlois, S.~Renaux-Petel, D.~A.~Steer and T.~Tanaka,
  ``Primordial fluctuations and non-Gaussianities in multi-field DBI
  inflation,''
  Phys.\ Rev.\ Lett.\  {\bf 101} (2008) 061301
  [arXiv:0804.3139 [hep-th]].
  
\bibitem{Langlois:2008qf}
  D.~Langlois, S.~Renaux-Petel, D.~A.~Steer and T.~Tanaka,
  ``Primordial perturbations and non-Gaussianities in DBI and general
  multi-field inflation,''
  Phys.\ Rev.\  D {\bf 78} (2008) 063523
  [arXiv:0806.0336 [hep-th]].
  
\bibitem{Langlois:2009ej}
  D.~Langlois, S.~Renaux-Petel and D.~A.~Steer,
  ``Multi-field DBI inflation: introducing bulk forms and revisiting the
  gravitational wave constraints,''
  JCAP {\bf 0904} (2009) 021
  [arXiv:0902.2941 [hep-th]].

\bibitem{RenauxPetel:2009sj}
  S.~Renaux-Petel,
  ``Combined local and equilateral non-Gaussianities from multifield DBI
  inflation,''
  JCAP {\bf 0910} (2009) 012
  [arXiv:0907.2476 [hep-th]].

 \bibitem{Achucarro:2010jv}
  A.~Achucarro, J.~O.~Gong, S.~Hardeman, G.~A.~Palma and S.~P.~Patil,
  ``Mass hierarchies and non-decoupling in multi-scalar field dynamics,''
  arXiv:1005.3848 [hep-th].

\bibitem{Kallosh:2004yh}
  R.~Kallosh and A.~D.~Linde,
  ``Landscape, the scale of SUSY breaking, and inflation,''
  JHEP {\bf 0412} (2004) 004
  [arXiv:hep-th/0411011].

\bibitem{Senatore:2010wk}
  L.~Senatore and M.~Zaldarriaga,
  ``The Effective Field Theory of Multifield Inflation,''
  arXiv:1009.2093 [hep-th].

\bibitem{Achucarro:2010da}
  A.~Achucarro, J.~O.~Gong, S.~Hardeman, G.~A.~Palma and S.~P.~Patil,
  ``Features of heavy physics in the CMB power spectrum,''
  arXiv:1010.3693 [hep-ph].
  



\end{thebibliography}
\end{document}